\documentclass[aps,prxquantum,twocolumn,notitlepage,superscriptaddress]{revtex4-1}
\usepackage{amsmath,amssymb}	
\usepackage{natbib}
\usepackage{url}
\usepackage{bm} 
\usepackage{mathtools}
\usepackage{braket}
\usepackage{hyperref}
\usepackage{graphicx}
\usepackage{color}
\usepackage{xcolor}
\usepackage[caption=false]{subfig}
\usepackage{epstopdf}
\usepackage{array}
\usepackage{empheq}
\usepackage{makecell}
\usepackage{booktabs}

\DeclareMathOperator{\Tr}{Tr}

\begin{document}
\title{Enhanced Quantum State Transfer and Bell State Generation over Long-Range Multimode Interconnects via Superadiabatic Transitionless Driving}

\author{Moein Malekakhlagh}
\affiliation{IBM Quantum, Thomas J. Watson Research Center, 1101 Kitchawan Rd, Yorktown Heights, NY, 10598, USA}
\author{Timothy Phung}
\affiliation{IBM Quantum, Almaden Research Center, San Jose, CA, 95120, USA}
\author{Daniel Puzzuoli}
\affiliation{IBM Quantum, IBM Canada, 750 West Pender St, Vancouver, BC, V6C 2T8, Canada}
\author{Kentaro Heya}
\affiliation{IBM Quantum, Thomas J. Watson Research Center, 1101 Kitchawan Rd, Yorktown Heights, NY, 10598, USA}
\author{Neereja Sundaresan}
\affiliation{IBM Quantum, Thomas J. Watson Research Center, 1101 Kitchawan Rd, Yorktown Heights, NY, 10598, USA}
\author{Jason Orcutt}
\affiliation{IBM Quantum, Thomas J. Watson Research Center, 1101 Kitchawan Rd, Yorktown Heights, NY, 10598, USA}

\begin{abstract}
Achieving high-fidelity direct two-qubit gates over meter-scale long quantum interconnects is challenging in part due to the multimode nature of such systems. One alternative scheme is to combine local operations with remote quantum state transfer or remote entanglement. Here, we study quantum state transfer and entanglement generation for two distant qubits, equipped with tunable interactions, over a common multimode interconnect. We employ the SuperAdiabatic Transitionless Driving (SATD) solutions for adiabatic passage and demonstrate various favorable improvements over the standard protocol. In particular, by suppressing leakage to a select (resonant) interconnect mode, SATD breaks the speed-limit relation imposed by the qubit-interconnect interaction $g$, where instead the operation time is limited by leakage to the adjacent modes, i.e. free spectral range $\Delta_c$ of the interconnect, allowing for fast operations even with weak $g$. Furthermore, we identify a multimode error mechanism for Bell state generation using such adiabatic protocols, in which the even/odd modal dependence of qubit-interconnect interaction breaks down the dark state symmetry, leading to detrimental adiabatic overlap with the odd modes growing as $(g/\Delta_c)^2$. Therefore, adopting a weak coupling, imposed by a multimode interconnect, SATD provides a significant improvement in terms of operation speed and consequently sensitivity to incoherent error.
\end{abstract}

\date{\today}
\maketitle

\section{Introduction}
\label{Sec:Intro}

Modular design of quantum computers \cite{Devitt_Architectural_2009, Monroe_Large_2014, Bravyi_Future_2022} relaxes wiring and control complexity, as well as cryogenic cooling power requirements, of the underlying Quantum Processing Units (QPU), and is the path forward for the required scaling \cite{Fowler_High_2009, Fowler_Surface_2012, Bravyi_High_2023} towards quantum error correction \cite{Knill_Theory_1997, Gottesman_Stabilizer_1997, Knill_Theory_2000, Lidar_Quantum_2013, Terhal_Quantum_2015}. For superconducting qubits, this vision necessitates developing interconnects at various levels of modularity \cite{Bravyi_Future_2022}, such as dense short-range interconnects \cite{Gold_Entanglement_2021, Conner_Superconducting_2021} to extend the effective size of QPUs, and sparse meter-range interconnects to enable parallelization of multiple QPUs within a dilution fridge. The short-range interconnect length is comparable to the distance between the qubits within a single chip and behaves effectively as a single mode system. While standard two-qubit gates could potentially work across a short interconnect, the multimode nature of long-range interconnects makes \textit{direct} two-qubit gates more difficult. Two alternatives are to use the interconnect to perform state transfer, or to generate remote entanglement such as a Bell state. In conjunction with local operations and classical communication, either of these operations can be used as a resource to implement indirect remote two-qubit gates \cite{Gottesman_Demonstrating_1999, Eisert_Optimal_2000, Huang_Experimental_2004}.  

\begin{figure}[h!]
\centering
\includegraphics[scale=0.335]{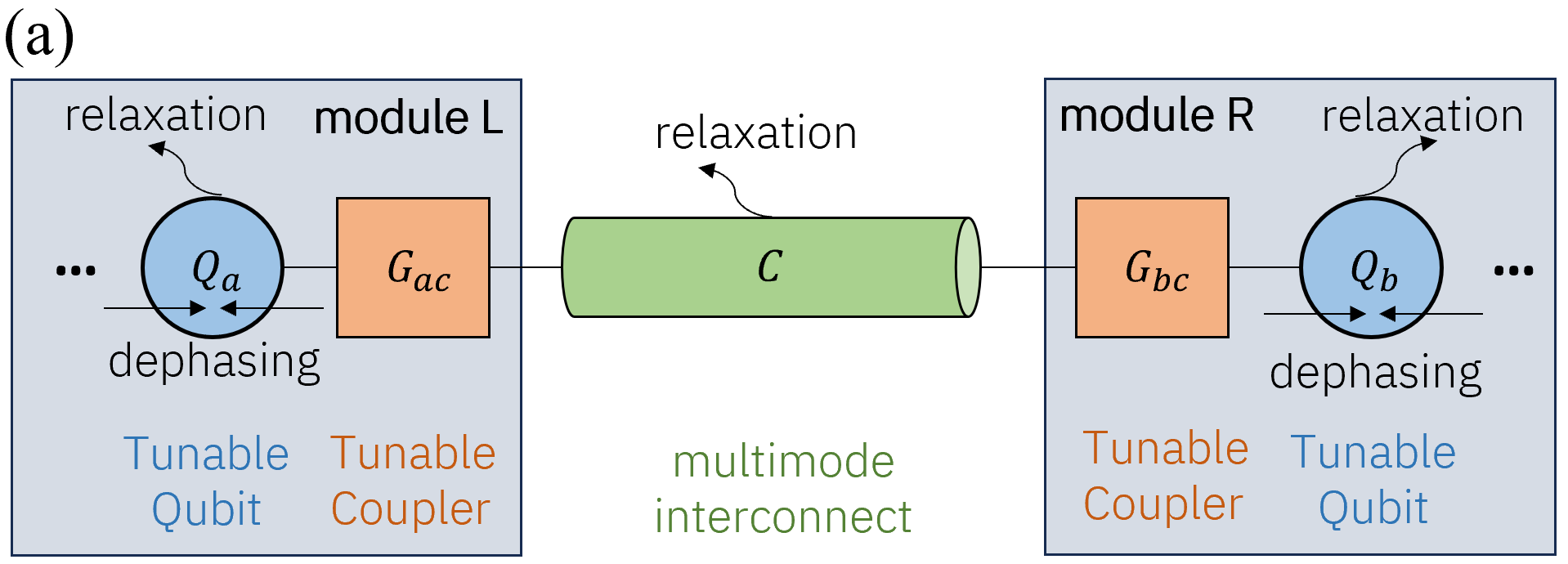} \\
\vspace{2mm}
\includegraphics[scale=0.445]{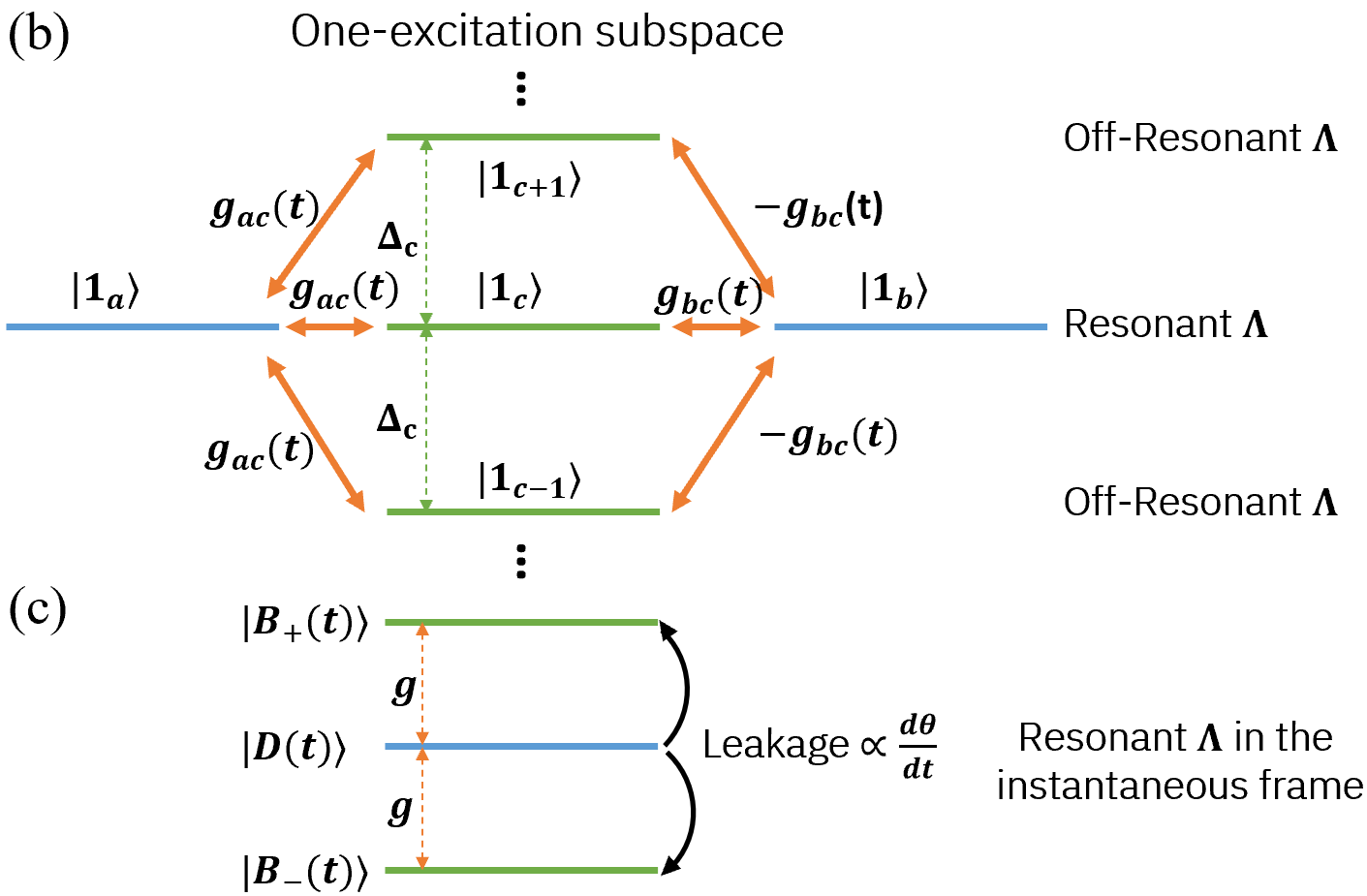} 
\caption{\textbf{System schematics and energy level diagram:} (a) Two quantum modules each with tunable-frequency qubits and tunable coupling to a shared multimode interconnect. We account for incoherent error due to qubit/interconnect relaxation and qubit pure dephasing. (b) Energy-level diagram in the one-excitation subspace. Bringing the qubits into resonance with a center mode forms a resonant Lambda system. There are, however, off-resonant Lambda systems formed by the adjacent interconnect modes with mode-dependent interaction sign that are detuned by FSR $\Delta_c$. (c) The resonant Lambda system in the instantaneous frame forms a dark eigenstate, using which one can implement STIRAP and its enhanced version SATD \cite{Baksic_Speeding_2016}, where we actively cancel out the dark-bright leakage transition. The leakage is proportional to the derivative of the STIRAP mixing angle $\theta(t)\equiv \arctan(g_{ac}(t)/g_{bc}(t))$ (Appendices~\ref{App:STIRAP} and~\ref{App:SATD}).}
\label{fig:system_schematics}
\end{figure}

The past few years have seen a recent surge into numerous superconducting circuit realizations of remote entanglement generation and quantum state transfer \cite{Kurpiers_Deterministic_2018, Axline_Demand_2018, Campagne_Deterministic_2018, Leung_Deterministic_2019, Zhong_Violating_2019, Bienfait_Phonon_2019, Chang_Remote_2020, Zhong_Deterministic_2021, Yan_entanglement_2022, Kannan_Demand_2023, Niu_Low_2023, Qiu_Deterministic_2023}. These protocols can be broadly categorized as either employing time-symmetric emission and capture of itinerant photons \cite{Kurpiers_Deterministic_2018, Axline_Demand_2018, Campagne_Deterministic_2018, Zhong_Violating_2019, Bienfait_Phonon_2019, Kannan_Demand_2023, Niu_Low_2023, Qiu_Deterministic_2023}, or using qubit interactions with the standing-wave modes of meter-long scale interconnects \cite{Leung_Deterministic_2019, Zhong_Violating_2019, Chang_Remote_2020, Zhong_Deterministic_2021, Qiu_Deterministic_2023}. Among protocols based on standing-wave modes, Stimulated Raman Adiabatic Passage (STIRAP) \cite{Gaubatz_Population_1990, Vitanov_Stimulated_2017, Bergmann_Roadmap_2019} achieves better fidelity \cite{Chang_Remote_2020} compared to a qubit-interconnect-qubit direct excitation exchange, also referred to as the relay protocol \cite{Zhong_Violating_2019}, by protection against interconnect loss.

In STIRAP \cite{Gaubatz_Population_1990, Vitanov_Stimulated_2017, Bergmann_Roadmap_2019}, we evolve the dark eigenstate of a Lambda system adiabatically towards a desired target state, applicable to quantum state transfer and entanglement generation. One advantage is the suppression of potential relaxation through the intermediate lossy interconnect. The operation speed is, however, limited by leakage to the bright lossy eigenstates whose transition frequency is set by the coupling strength. Transitionless Driving (TD) methods \cite{Demirplak_Adiabatic_2003, Demirplak2008consistency, Berry_Transitionless_2009, Ibanez_Multiple_2012, Zheng_Optimal_2022} cancel out non-adiabatic transitions \textit{exactly} via a modified control Hamiltonian, similar in spirit to the \textit{perturbative} Derivative Removal by Adiabatic Gate (DRAG) technique \cite{Motzoi_Simple_2009, Gambetta_Analytic_2011, Malekakhlagh_Mitigating_2022, Li_Suppression_2023}. One potential practical drawback can however be the need for a control knob not accessible by the original Hamiltonian. SuperAdiabatic Transition Driving (SATD) \cite{Baksic_Speeding_2016, Zhou_Accelerated_2017} redefines the evolution path, connecting the original source and target states, such that in the dressed frame the non-adiabatic transitions are canceled out exactly without the need for additional control knobs. SATD solutions for STIRAP have also been generalized to single-qubit tripod gates \cite{Ribeiro_Accelerated_2019, Setiawan_Analytic_2021} and more recently to two-qubit gates \cite{Setiawan_Fast_2023} for fluxonium qubits \cite{Manucharyan_Fluxonium_2009}.  

In this paper, we characterize the performance improvements of the SATD protocol against STIRAP, and promote its usage for quantum state transfer and Bell state generation in a \textit{multimode} interconnect setting. By removing leakage to the resonant interconnect mode, the operation speed for SATD is not limited by the qubit-interconnect coupling $g$ anymore, but determined by the interconnect Free Spectral Range (FSR) $\Delta_c$, leading to a significant speedup as well as a robustness to variation in $g$ compared to STIRAP. We show that the single-mode SATD solutions work reasonably well for a multimode interconnect with sufficiently large FSR ($\Delta_c \gg g$), and quantify the deviations from expected behavior due to multimode effects. In particular, we find that the even-odd mode dependence of the interaction breaks the dark-state symmetry, which is in principle detrimental to such dark-state-based adiabatic protocols. This impacts the Bell state generation more by an adiabatic overlap error proportional to $(g/\Delta_c)^2$, and can be mitigated only via a weaker $g$. This weaker $g$ requirement due to multimode effects, and the $g$ robustness of SATD makes its application very advantageous especially for Bell state generation. Furthermore, we observe improvements by SATD in suppressing the incoherent error due to qubits relaxation, pure dephasing, and the interconnect quality factor.    
 
The remainder of this work is organized as follows. Section~\ref{Sec:model} describes the system under consideration with two quantum modules connected via a multimode interconnect, and a Lindblad model introduced for our analytical and numerical analyses. In Sec.~\ref{Sec:STIRAP}, we revisit the ideal single-mode STIRAP protocol, used for quantum state transfer and entanglement generation, and discuss potential detrimental multimode sources of error. In Sec.~\ref{Sec:SATD}, we present extensive simulations investigating the numerous advantages of SATD compared to regular STIRAP in such a multimode context. We further assess the performance of indirect two-qubit gates achieved by combining quantum state transfer and remote entanglement with local operations. Appendix~\ref{App:Lindblad} discusses the details of our Lindblad model and simulations, and provides a numerical convergence test. Appendices~\ref{App:STIRAP} and~\ref{App:MMSTIRAP} review the single-mode STIRAP, and complications that arise due to a multimode interconnect, respectively. In Appendix~\ref{App:SATD}, we review the derivation of SATD solutions for single-mode STIRAP following Ref.~\cite{Baksic_Speeding_2016}.             

\section{system and model}
\label{Sec:model}

We consider a system consisting of two tunable-frequency qubits that have tunable interactions to a common long-range multimode interconnect as depicted in Fig.~\ref{fig:system_schematics}(a). The standard motivation for such a setup is to perform remote quantum operations between two modules (chips) connected via a long interconnect such as a coaxial cable. This can, however, be also relevant to on-chip transmission lines between distant qubits \cite{Sundaresan_Beyond_2015}. Notable experimental studies have employed tunable grounded Transmon qubits with tunable RF SQUID couplers \cite{Barends_Coherent_2013, Chen_Qubit_2014, Geller_Tunable_2015} connected via an on-chip transmission line \cite{Chang_Remote_2020} or a cable \cite{Qiu_Deterministic_2023}. Although this work is motivated by superconducting architectures, the following analysis and characterization of remote operations is presented in a system-agnostic manner.

\begin{figure*}[t!]
\centering
\includegraphics[scale=0.245]{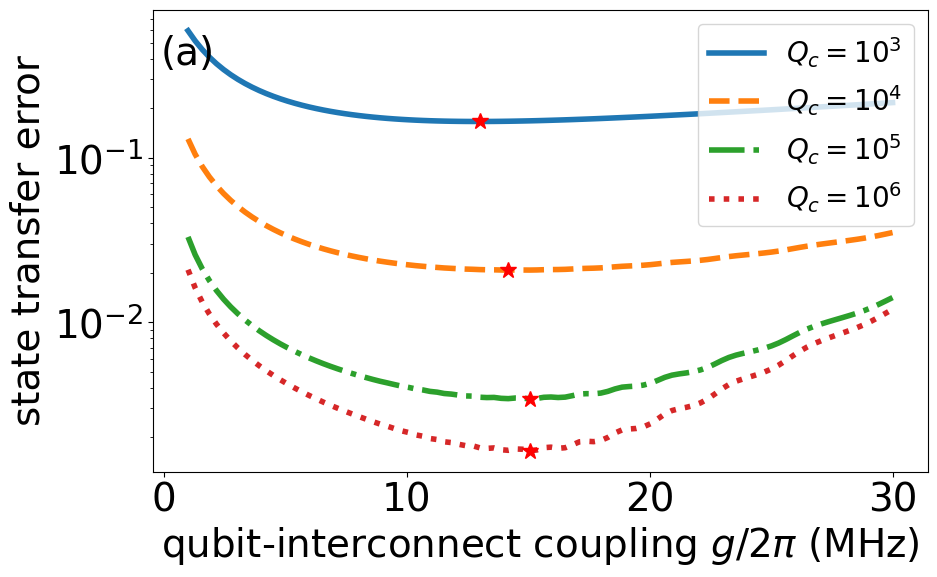}
\includegraphics[scale=0.245]{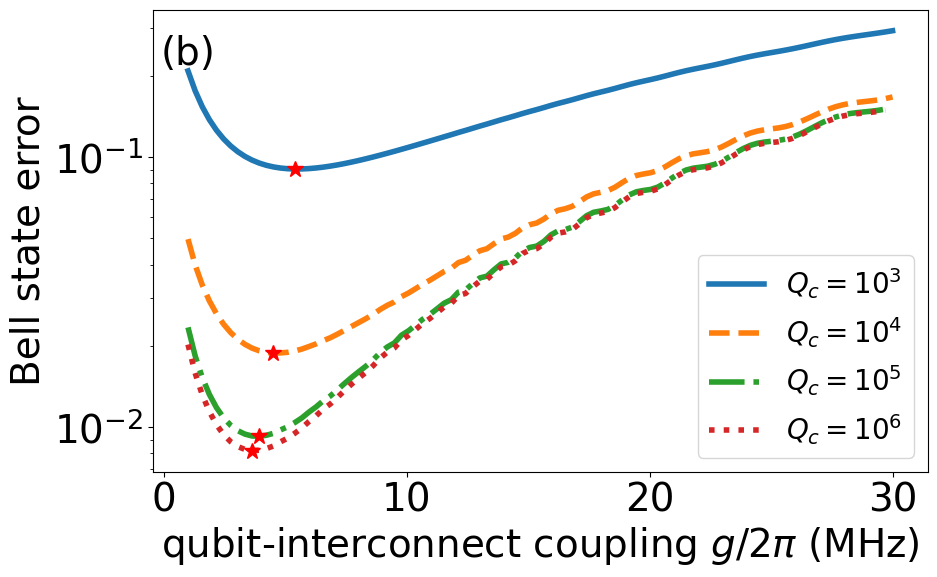}
\includegraphics[scale=0.245]{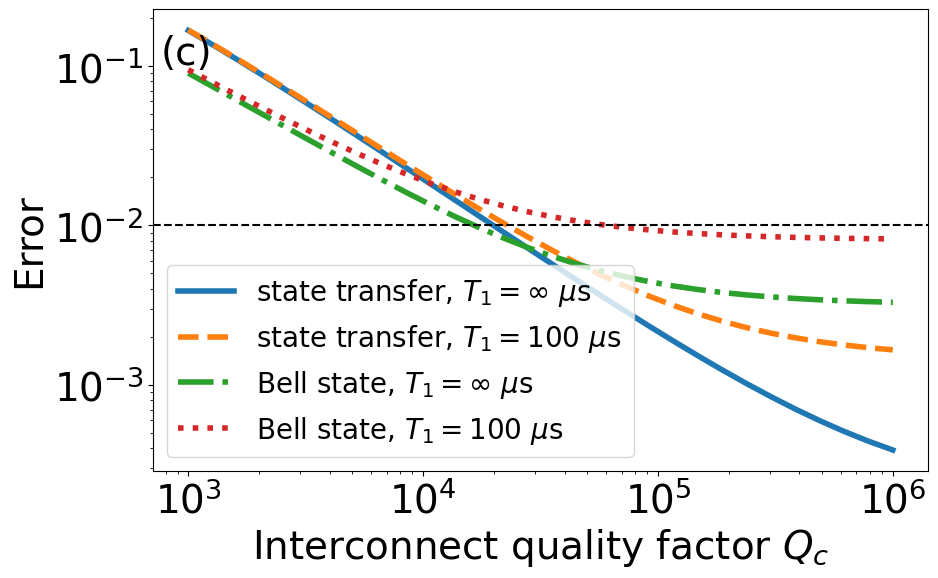}
\caption{\textbf{Characterization of STIRAP performance for state transfer and Bell state generation via a multimode interconnect:} (a) state transfer error, and (b) Bell state error, as a function of qubit-interconnect coupling $g$ for various interconnect quality factors $Q_c\in [10^3,10^6]$ (same for all modes). We included five interconnect modes, with $\Delta_c/2\pi =100$ MHz, where the center mode is resonant with the qubits (see Appendix~\ref{App:Lindblad} for a convergence test). Qubits relaxation is set to $T_{1,a}=T_{1,b}=100 \ \mu$s. Here, for each value of $g$, the operation time is set to minimize dark$\rightarrow$bright leakage as $g\tau_p = 4\pi$ \cite{Chang_Remote_2020}. STIRAP angles for state transfer and Bell state generation are $\theta_p = \pi/2$ and $\theta_p = \pi/4$, respectively. Note that optimal couplings for state transfer and Bell state (red stars) are distinct and are approximately found as $g/2\pi \approx 15$ and $4$ MHz, respectively. (c) State transfer and Bell state generation error as a function of $Q_c$ using the optimal $g$ in (a)--(b).}
\label{fig:MMSTIRAP-ST Bell Error}
\end{figure*}

We characterize the performance of STIRAP and SATD protocols for quantum state transfer and Bell state generation via a Lindblad simulation that accounts for qubit relaxation ($T_{1}$), pure dephasing ($T_{2\phi}$) and cable relaxation ($\kappa_n$):
\begin{align}
\begin{split}
\dot{\hat{\rho}}_s(t) &= - i [\hat{H}_s(t),\hat{\rho}_s(t)] + \sum\limits_{q=a,b} \frac{1}{T_{1q}}\mathcal{D}[\hat{q}]\hat{\rho}_s(t) \\
& + \sum\limits_{q=a,b} \frac{2}{T_{2\phi,q}}\mathcal{D}[\hat{q}^{\dag}\hat{q}]\hat{\rho}_s(t) + \sum\limits_{n} \kappa_n \mathcal{D}[\hat{c}_n]\hat{\rho}_s(t) \;,
\end{split} 
\label{eq:model-Lindblad eq.}
\end{align}
where $\hat{\rho}_s(t)$ is the system density matrix, and $\mathcal{D}[\hat{C}]\hat{\rho}_s \equiv \hat{C}\hat{\rho}_s\hat{C}^{\dag} - (1/2)\{\hat{C}^{\dag}\hat{C},\hat{\rho}_s\}$ is the dissipator for the collapse operator $\hat{C}$. We model the qubits as weakly anharmonic, and the interconnect as a collection of harmonic quantum oscillators, with time-dependent (controllable) qubit-interconnect interaction as:  
\begin{align}
\begin{split}
\hat{H}_s(t) & = \sum\limits_{q=a,b}\left[\omega_q(t) \hat{q}^{\dag}\hat{q}+\frac{\alpha_q}{2} \hat{q}^{\dag}\hat{q}^{\dag}\hat{q}\hat{q}\right] + \sum\limits_{n-n_c=-N}^{N} \omega_{n} c_n^{\dag}\hat{c}_n \\
&+ \sum\limits_{n-n_c=-N}^{N} g_{an}(t) \left(\hat{a}\hat{c}_n^{\dag}+\hat{a}^{\dag}\hat{c}_n\right) \\
&+ \sum\limits_{n-n_c=-N}^{N} g_{bn}(t) (-1)^n\left(\hat{b}\hat{c}_n^{\dag}+\hat{b}^{\dag}\hat{c}_n\right) \;,
\end{split} 
\label{eq:model-Def of H(t)}
\end{align}
where $\omega_q(t)$, $\alpha_q$, $\omega_{n}=\omega_{n_c} + n \Delta_c$ and $g_{qn}(t)$ are the qubit frequency, anharmonicity, evenly spaced $n^{th}$ mode frequency with FSR $\Delta_c$, and qubit-interconnect interaction rates for $q=a,b$, respectively. Moreover, $n_c$ is the center mode index, and $N$ is the additional modes kept on each side. In writing Hamiltonian~(\ref{eq:model-Def of H(t)}), we have made certain approximations, motivated by the physics of multimode interconnects (see Appendix~\ref{App:Lindblad}), similar to Ref.~\cite{Chang_Remote_2020}. An important feature of Hamiltonian~(2) is the even-odd mode-dependent relative sign for the qubit-interconnect interaction, which accounts for the distinct spatial profile of even and odd interconnect modes \cite{Chang_Remote_2020}.  

\section{STIRAP via a multimode interconnect}
\label{Sec:STIRAP}

STIRAP is a protocol for adiabatic transfer of population in a Lambda system, i.e. between two quantum states coupled through a common intermediate state, via temporal control of the interactions [Fig.~\ref{fig:system_schematics}(b)]. Under the single-mode (ideal) case, the Hamiltonian reads
\begin{align}
\begin{split}
\hat{H}_{\text{STRP}}(t) & = \begin{bmatrix}
0 & g_{ac}(t) & 0\\
g_{ac}(t) & 0 & g_{bc}(t) \\
0 & g_{bc}(t) & 0 
\end{bmatrix} \;,
\end{split}
\label{eq:STIRAP-Def of H_strp}
\end{align} 
where we assume all levels are resonant. This resonant Lambda system has a dark eigenstate
\begin{align}
\ket{D(t)} \equiv \cos\theta(t) \ket{1_a0_c0_b}-\sin\theta(t)  \ket{0_a0_c1_b} \;,
\end{align}
and two bright eigenstates 
\begin{align}
\begin{split}
\ket{B_{\pm}(t)} \equiv \frac{1}{\sqrt{2}} &[  \sin\theta(t) \ket{1_a0_c0_b} \pm \ket{0_a1_c0_b} \\
&-\cos\theta(t)  \ket{0_a0_c1_b}] \;,
\end{split}
\end{align}
with the mixing angle defined as $\tan\theta(t)\equiv g_{ac}(t)/g_{bc}(t)$.

The dark eigenstate, having no overlap with the intermediate (possibly) lossy interconnect state, therefore allows for an adiabatic quantum state transfer by arbitrarily evolving the mixing angle $\theta(t)$. A common choice for the controls are $g_{ac}(t) =g\sin\theta(t)$ and $g_{bc}(t) =g\cos\theta(t)$ with $\theta(t) =  (t/\tau_p)\theta_p$ for $t\in[0,\tau_p]$. For instance, sweeping $\theta(t)$ from 0 to $\pi/2$ or $\pi/4$ should ideally implement $\ket{1_a0_c0_b}\rightarrow -\ket{0_a0_c1_b}$ (full state transfer), or $\ket{1_a0_c0_b}\rightarrow 1/\sqrt{2}(\ket{1_a0_c0_b}-\ket{0_a0_c1_b})$ (Bell state), respectively. Note, however, that unwanted non-adiabatic $\ket{D(t)}\rightarrow\ket{B_{\pm}(t)}$ transitions, whose probabilities grow with $\dot{\theta}(t)$, impose a limit on the STIRAP speed (Fig.~\ref{fig:system_schematics}(c) and Appendix~\ref{App:STIRAP}).

For a multimode interconnect, with the interaction $g$ comparable to the FSR $\Delta_c$, the adjacent modes impact the fidelity of STIRAP detrimentally by (i) breaking the dark-symmetry condition, (ii) introducing additional leakage, and (iii) additional decay channels. Regarding item (i), each adjacent mode forms an effective off-resonant Lambda system with the qubits [Fig.~\ref{fig:system_schematics}(b)]. Our extended multimode STIRAP analysis suggests that for a hypothetical interconnect with same-sign interactions the Hamiltonian supports the original dark eigenstate, while for the physical case of mode-dependent interactions, one instead finds a pseudo dark eigenstate having a non-zero overlap $|[g(t)/\Delta_c]\sin[2\theta(t)]|$ with the one-excitation subspace of the odd (opposite-sign) modes (Appendix~\ref{App:MMSTIRAP}). Such an adiabatic error vanishes for quantum state transfer with $\theta(\tau_p)=\pi/2$, but is maximized for Bell state generation with $\theta(\tau_p)=\pi/4$ requiring a weaker $g$ for better fidelity.  

\begin{figure*}[t!]
\centering
\includegraphics[scale=0.245]{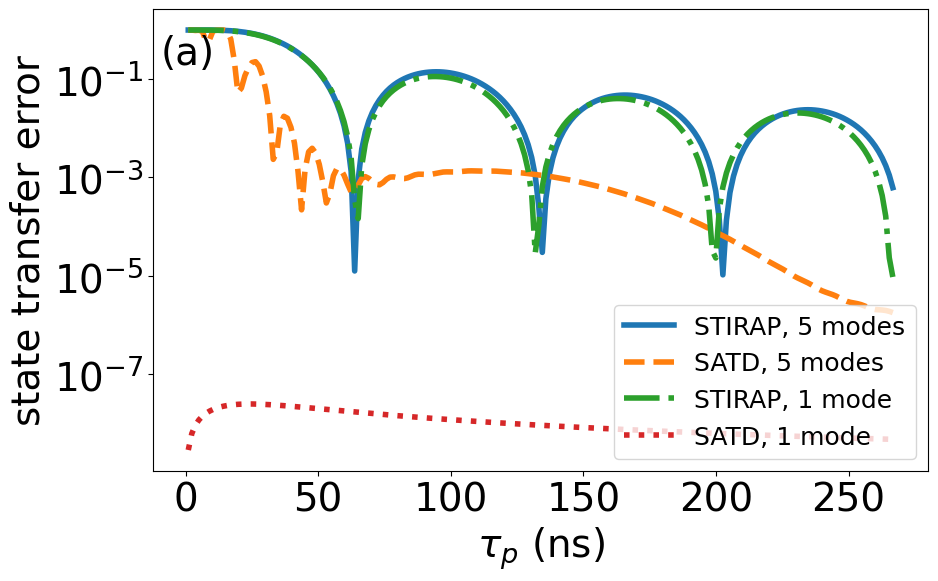}
\includegraphics[scale=0.245]{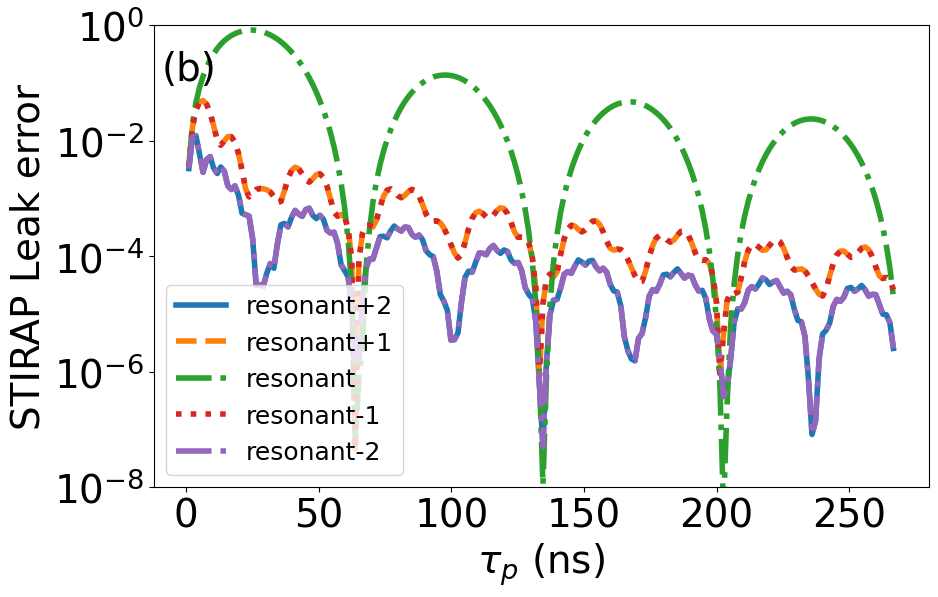}
\includegraphics[scale=0.245]{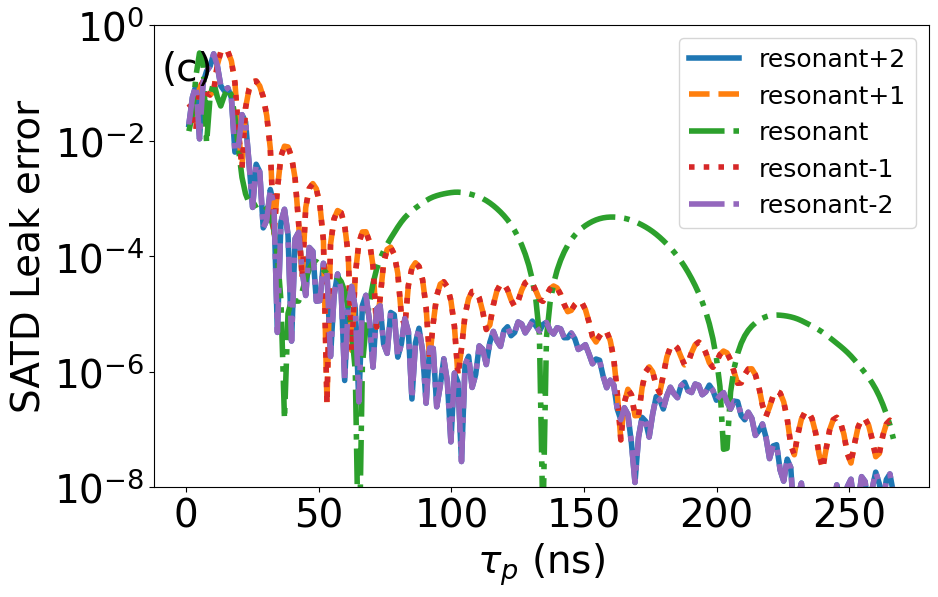}
\caption{\textbf{STIRAP versus SATD performance comparison and leakage breakdown for quantum state transfer}: (a) state-transfer error considering one (ideal) and five interconnect modes, (b)--(c) final leakage, i.e. at $t=\tau_p$, to the interconnect modes for the five-mode STIRAP and SATD simulations, respectively. The result is found by numerical simulation of Eqs.~(\ref{eq:model-Lindblad eq.})--(\ref{eq:model-Def of H(t)}), where here the incoherent relaxation and dephasing channels are turned off. Qubit-interconnect interactions and FSR are set to $g/2\pi = 15$ (optimal choice from Fig.~\ref{fig:MMSTIRAP-ST Bell Error}) and $\Delta_c/2\pi=100$ MHz, respectively.}
\label{fig:SATD-STIRAP-SATD ST comparison}
\end{figure*}

Figure~\ref{fig:MMSTIRAP-ST Bell Error} characterizes the performance of STIRAP in such a multimode context, where we simulate the Lindblad Eqs.~(\ref{eq:model-Lindblad eq.})--(\ref{eq:model-Def of H(t)}) numerically with five interconnect modes evenly spaced about the qubit frequency for the common sine/cosine STIRAP controls and initial pure state $\hat{\rho}_s(0)\equiv \ket{1_a0_{c}0_b} \bra{1_a0_{c}0_b}$. We define error,
\begin{align}
E \equiv 1- \Tr\{\hat{\rho}_s(\tau_p)\ket{\psi_{\text{id}}}\bra{\psi_{\text{id}}}\} \;,
\end{align}
in terms of the overlap of the final density matrix with the ideal target states for state transfer and Bell states as $\ket{\psi_{\text{id}}}=\ket{0_a0_c1_b}$ and $\ket{\psi_{\text{id}}}=(1/\sqrt{2})(\ket{1_a0_c0_b}-\ket{0_a0_c1_b})$, respectively. Panels (a) and (b) show the corresponding error as a function of $g$ for fixed FSR of $\Delta_c/2\pi=100$ MHz and various interconnect quality factors, where we observe distinct approximate optimal $g/\Delta_c$ ratio of $15\%$ and $4\%$ for state transfer and Bell state, respectively, at $Q_c=10^5$ and $T_{1a}=T_{1b}=100 \ \mu$s. The optimal couplings are a balance between more leakage to the neighboring modes at stronger $g$ (faster operation) and more incoherent error at weaker $g$ (slower operation). The Bell state generation also suffers from a non-zero \textit{adiabatic} overlap with the odd interconnect modes due to dark-state symmetry breakdown (Appendix~\ref{App:MMSTIRAP} and Sec.~\ref{SubSec:SATD-Bell}). Using the optimal couplings in panel (c), we find that in order to reach sub-percent error the interconnect $Q_c$ must approximately exceed $2.2\times 10^{4}$ and $6.5\times 10^{4}$ for state transfer and Bell state, respectively.   

\begin{figure}[t!]
\centering
\includegraphics[scale=0.37]{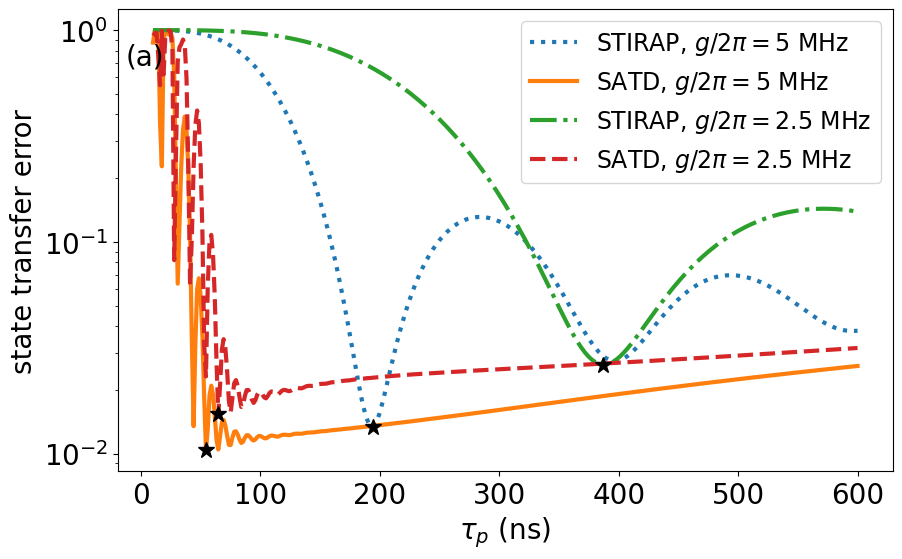} \\
\includegraphics[scale=0.37]{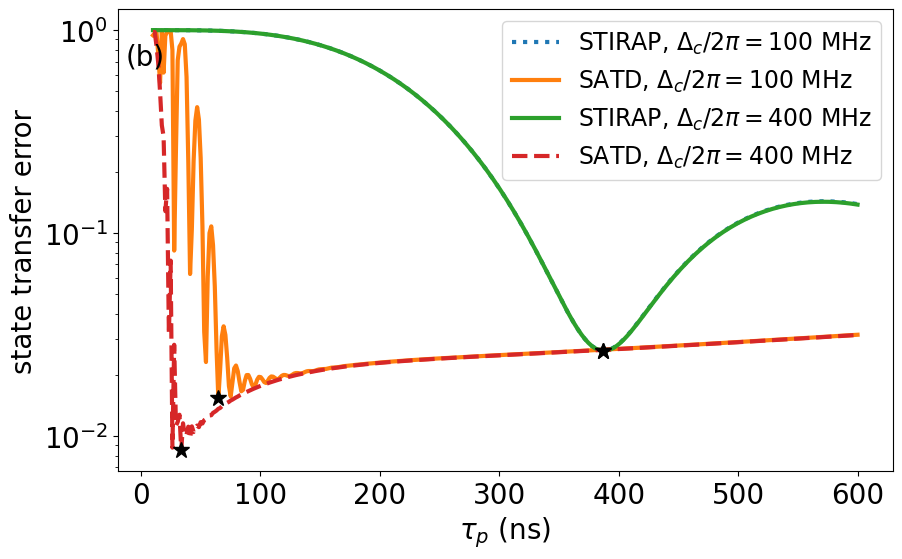} 
\caption{\textbf{Robustness of SATD with respect to qubit-interconnect interaction $g$:} state transfer error as a function of $\tau_p$ for (a) two \textit{weak} values of $g/2\pi=5$ and $2.5$ MHz, (b) two values of FSR $\Delta_c/2\pi = 100$ and $400$ MHz and fixed $g/2\pi=2.5$ MHz. Other parameters are set as $T_{1,a}=T_{1,b}=100 \ \mu$s, $T_{2\phi,a}=T_{2\phi,b}=10 \ \mu$s, $Q_c=10^{5}$. These comparisons emphasize the distinct error behavior, where for STIRAP it is mostly dependent on the choice of $g$, and for SATD it is limited by the FSR/length of the interconnect. The two STIRAP curves in panel (b) lie on top.}
\label{fig:SATD-g robustness}
\end{figure}

\section{Improved STIRAP via SuperAdiabatic Transitionless Driving}
\label{Sec:SATD}

TD is a control technique for cancelling out non-adiabatic transitions via a corrected control Hamiltonian \cite{Demirplak_Adiabatic_2003, Demirplak2008consistency, Berry_Transitionless_2009, Ibanez_Multiple_2012}. In an ideal case, from the hardware perspective, the correction can be simply implemented via a modification of the original control pulses. The superadiabatic aspect refers to implementing the cancellation in a dressed frame, i.e. effectively evolving the initial state in a modified path in the Hilbert space towards the target state. The standard single-mode STIRAP problem allows for a family of exact SATD solutions \cite{Baksic_Speeding_2016}. A commonly employed SATD solution dresses the evolution path along the spin-1 $\hat{M}_x$ operator, yielding the explicit results \cite{Baksic_Speeding_2016, Ribeiro_Accelerated_2019, Setiawan_Analytic_2021, Setiawan_Fast_2023} (see also Appendix~\ref{App:SATD})
\begin{align}
& g_{ac}(t) = g \Big[\sin\theta(t) +\frac{\cos [\theta(t)] \ddot{\theta} (t)}{g^2+\dot{\theta}(t)^2}\Big] \;,
\label{eq:SATD-Def of gac}\\
& g_{bc}(t) = g \Big[\cos\theta(t) -\frac{\sin [\theta(t)] \ddot{\theta} (t)}{g^2+\dot{\theta}(t)^2}\Big] \;,
\label{eq:SATD-Def of gbc}
\end{align}
where the corrections depend on both the first $\dot{\theta}(t)$ and the second derivative $\ddot{\theta}(t)$ of the mixing angle (check Appendix~\ref{App:SATD} for a comparison of pulse shapes).    

\begin{figure}[t!]
\centering
\includegraphics[scale=0.37]{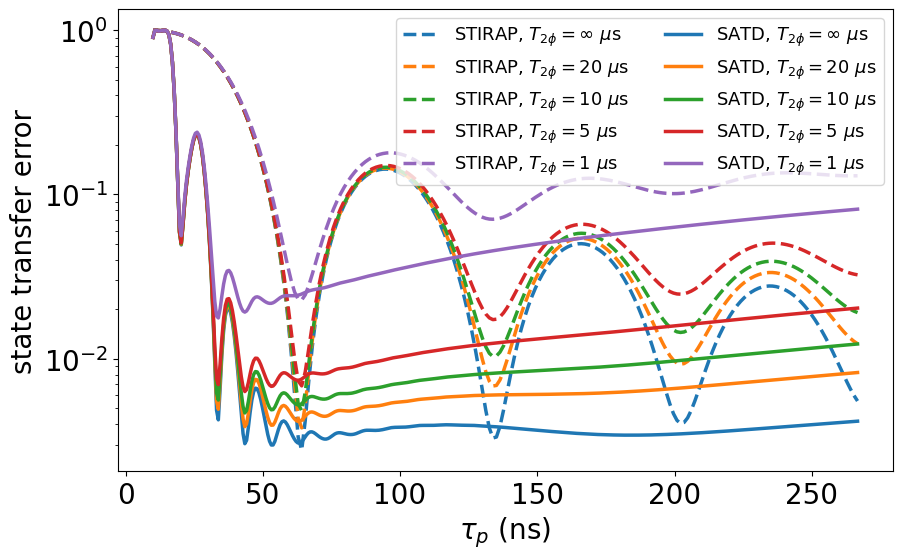} 
\caption{\textbf{Dependence of state transfer error on pure dephasing:} comparing STIRAP and SATD as a function of $\tau_p$ for various dephasing times $T_{2\phi,a}=T_{2\phi,a} \in \{\infty, 20, 10, 5, 1\} \ \mu$s. Other parameters are set as $T_{1a}=T_{1b}=100 \ \mu$s, $Q_c=10^{5}$, $g/2\pi=15$  MHz, and $\Delta_c/2\pi=100$ MHz. STIRAP and SATD results are shown with dashed and solid curves with the same colors for each $T_{2\phi}$ value.}
\label{fig:SATD-ST deph}
\end{figure}

\begin{figure}[t!]
\centering
\includegraphics[scale=0.37]{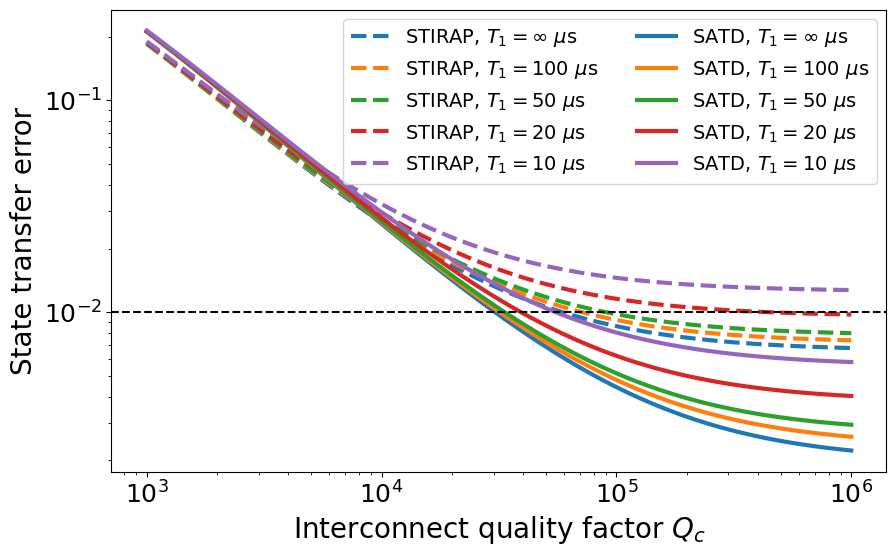} 
\caption{\textbf{Dependence of state transfer error on qubit/interconnect relaxation:} comparing STIRAP (dashed) and SATD (solid) results as a function of interconnect quality factor $Q_c$ for various qubit lifetimes of $T_{1a}=T_{1b}\in \{\infty, 100, 50, 20, 10\}$ $\mu$s, fixed $T_{2\phi,a}=T_{2\phi,b}=10$ $\mu$s and $g/2\pi=15$ MHz. Operation times for STIRAP (SATD) are picked according to the corresponding minimum in Fig.~\ref{fig:SATD-ST deph} as $\tau_p=65$ ns ($\tau_p=44$ ns).}
\label{fig:SATD-ST Relax}
\end{figure}

\begin{figure*}[t!]
\centering
\includegraphics[scale=0.245]{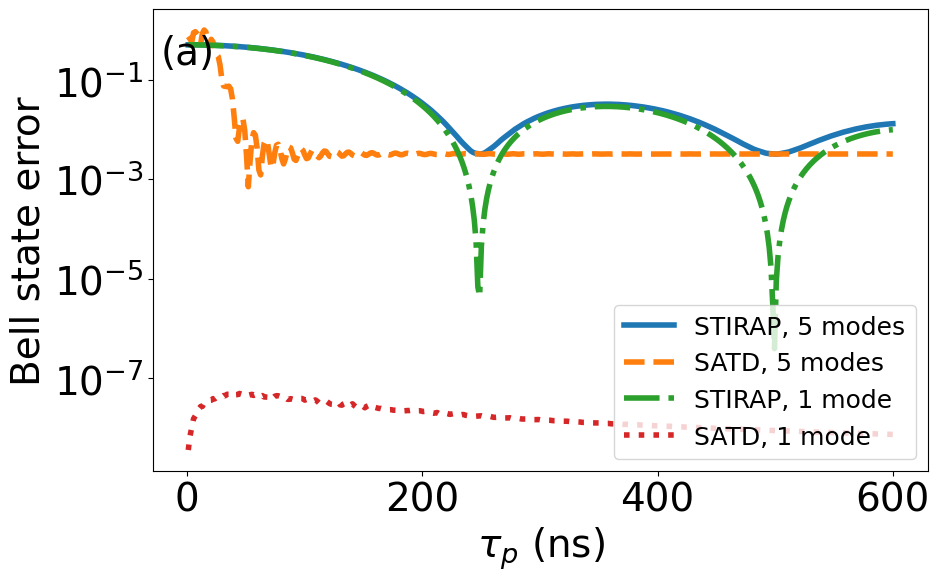}
\includegraphics[scale=0.244]{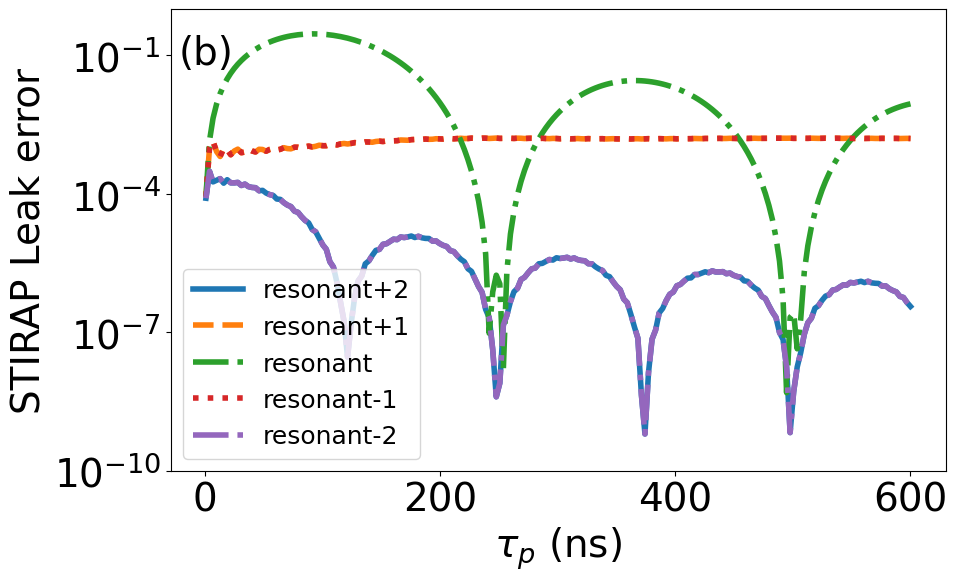}
\includegraphics[scale=0.244]{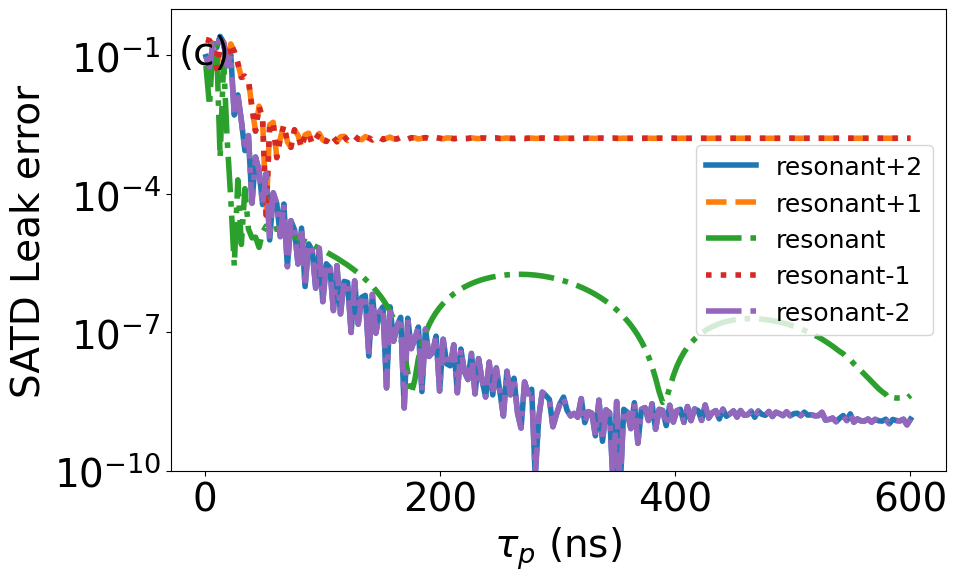}
\caption{\textbf{STIRAP versus SATD performance comparison and leakage breakdown for Bell state generation}: This figure has the same format as Fig.~\ref{fig:SATD-STIRAP-SATD ST comparison} except for a weaker qubit-interconnect $g/2\pi = 4$ MHz (approximately optimal based on Fig.~\ref{fig:MMSTIRAP-ST Bell Error}(b)). The enhanced population of the odd interconnect modes is caused by the even/odd coupling sign, and is more detrimental to the Bell state generation.}
\label{fig:SATD-STIRAP-SATD BELL comparison}
\end{figure*}

In the following, we characterize and compare the performance of regular STIRAP and SATD protocols, and dissect various favorable aspects of SATD usage in the context of multimode interconnects for quantum state transfer and Bell state generation. Some advantages of the SATD protocol can be summarized as follows: 
\begin{itemize}
\item[(i)] The speed of standard STIRAP is limited by the dark$\rightarrow$bright transitions, whose effective transition frequency is equivalent to the resonant interaction rate $g$ (Appendix~\ref{App:STIRAP}). SATD, however, removes the dark$\rightarrow$bright leakage, and allows for faster operations whose speed limit is set by leakage to the adjacent interconnect modes (Appendix~\ref{App:SATD}, Figs.~\ref{fig:SATD-STIRAP-SATD ST comparison} and~\ref{fig:SATD-STIRAP-SATD BELL comparison}). 

\item [(ii)] One crucial practical consequence of (i) is the SATD robustness to qubit-interconnect interaction $g$, and the possibility of performing fast high-fidelity operations even with weak interactions (Fig.~\ref{fig:SATD-g robustness}).

\item [(iii)] SATD provides a more pronounced speedup over STIRAP for the Bell state generation. This is the case as Bell state generation is more sensitive to the even/odd sign dependence of the interaction, compared to state transfer, and requires a weaker coupling to mitigate adiabatic overlap with the odd modes (Appendix~\ref{App:MMSTIRAP} and Fig.~\ref{fig:SATD-STIRAP-SATD BELL comparison}).    

\item [(iv)] We also observe improved sensitivity of the SATD protocol error, compared to STIRAP, with respect to qubit and interconnect mode relaxation, as well as qubit pure dephasing (Figs.~\ref{fig:SATD-ST deph}--\ref{fig:SATD-ST Relax} and \ref{fig:SATD-Bell deph}--\ref{fig:SATD-Bell Relax}). 
\end{itemize}

\subsection{Quantum state transfer}
\label{SubSec:SATD-QST}
    
We begin by analyzing the performance of SATD for quantum state transfer. Figure~\ref{fig:SATD-STIRAP-SATD ST comparison} shows a comparison and breakdown of quantum state transfer error between regular STIRAP and SATD protocols. To emphasize the corrections provided by adopting the single-mode SATD solutions in this multimode setting, here we present the results considering both single and five interconnect modes and for zero qubit pure dephasing and qubit/interconnect relaxation. Panel (a) shows that for regular STIRAP, the single- and five-mode curves agree indicating that state-transfer error is limited \textit{mainly} by leakage to the resonant interconnect mode. The single-mode SATD simulation confirms the elimination of this leakage down to numerical error as expected. Applying the single-mode SATD solutions~(\ref{eq:SATD-Def of gac})--(\ref{eq:SATD-Def of gbc}) in the multimode setting is still advantageous in terms of operation time (orange curve). Panels (b)--(c) show the breakdown of final-time leakage to individual interconnect modes, where for SATD the error is mainly limited by leakage to the adjacent interconnect modes at shorter times. 

To demonstrate the robustness of the SATD protocol with respect to $g$, and the interplay with FSR $\Delta_c$, we compare STIRAP and SATD performance for weaker-than-optimal couplings in Fig.~\ref{fig:SATD-g robustness}. Comparing the state transfer error for fixed $\Delta_{c}/2\pi=100$ MHz and two choices of $g/2\pi=2.5$ and $5$ MHz in panel (a) reveals a $g$ robustness of the SATD protocol, in which the two couplings provide comparable optimal error (black stars) of 0.015 and 0.01 at 64 ns and 54 ns, respectively. On the other hand, the standard STIRAP's time is inversely proportional to $g$, where we find optimal error of approximately 0.013 and 0.026 at 194 ns and 387 ns (double), respectively. This feature of SATD is very beneficial as it allows to run high-fidelity fast operations even at weak coupling rates. In panel (b), we run a similar comparison but for a fixed weak $g/2\pi=2.5$ MHz and two interconnect FSR (inversely proportional to length) choices of $\Delta_c/2\pi = 100$ and $400$ MHz. By the same token, SATD provides faster and improved error for the larger FSR case, 0.009 at 34 ns compared to 0.015 at 65 ns, while regular STIRAP's error/speed is 0.026 at 387 ns and the same for the two cases.              

We also assess the improvement by SATD, compared to STIRAP, in error sensitivity to pure dephasing and qubit/interconnect relaxation in Figs.~\ref{fig:SATD-ST deph}--\ref{fig:SATD-ST Relax} with the approximately optimal $g/2\pi=15$ MHz for state transfer (Fig.~\ref{fig:MMSTIRAP-ST Bell Error}(a)). We find that the incoherent errors due to dephasing and relaxation are approximately additive. Figure~\ref{fig:SATD-ST deph} shows the state transfer error as a function of $\tau_p$ for various $T_{2\phi}$ ranging in $\in[\infty, 1] \ \mu$s. First, due to the expedited transfer, the pure dephasing error, $\Delta E_{\text{deph}}(T_{2\phi}) \equiv E_{T_{2\phi}}-E_{T_{2\phi}\rightarrow\infty}$, for SATD is substantially reduced, where at $\tau_p \approx 44$ ns we find $\Delta E_{\text{deph}}^{\text{SATD}}(1 \ \mu s)\approx 1.6\times 10^{-2}$. Moreover, at longer (standard) STIRAP time of $\tau_p = 4\pi/g \approx 130$ ns, SATD demonstrates a substantial improvement of dephasing error as $\Delta E_{\text{deph}}^{\text{SATD}}(1 \ \mu s)\approx 2.2\times 10^{-2}$ compared to $\Delta E_{\text{deph}}^{\text{STIRAP}}(1 \ \mu s)\approx 6.0\times 10^{-2}$. Furthermore, Fig.~\ref{fig:SATD-ST Relax} shows the error as a function of the quality factor $Q_c\in\{10^{3}, 10^{6}\}$ for various relaxation times $T_1\in \{\infty, 10\} \ \mu$s. We observe that SATD offers substantial improvement in sensitivity with respect to qubit $T_1$ as $\Delta E_{\text{q-rel}}^{\text{SATD}} (10 \ \mu s)\approx 3.6\times 10^{-3}$ compared to $\Delta E_{\text{q-rel}}^{\text{STIRAP}}(10 \ \mu s)\approx 6.0\times 10^{-3}$ at sufficiently large $Q_c$.       

\subsection{Bell state generation}
\label{SubSec:SATD-Bell}

We next discuss the advantages of the SATD protocol for Bell state (entanglement) generation. The trade-offs/benefits demonstrated in Sec.~\ref{SubSec:SATD-QST} for state transfer carry on to Bell state generation as well. In addition, Bell state generation is more prone to the even/odd sign of interaction, and hence requires a weaker qubit-interconnect $g$ as found in Fig.~\ref{fig:MMSTIRAP-ST Bell Error}(b). The use of regular STIRAP, however, means the operation time will be set by $g$ and hence would become very slow. Therefore, SATD, whose speed is limited by $\Delta_c$, provides a larger speedup for Bell state generation compared to state transfer.
       
\begin{figure}[t!]
\centering
\includegraphics[scale=0.37]{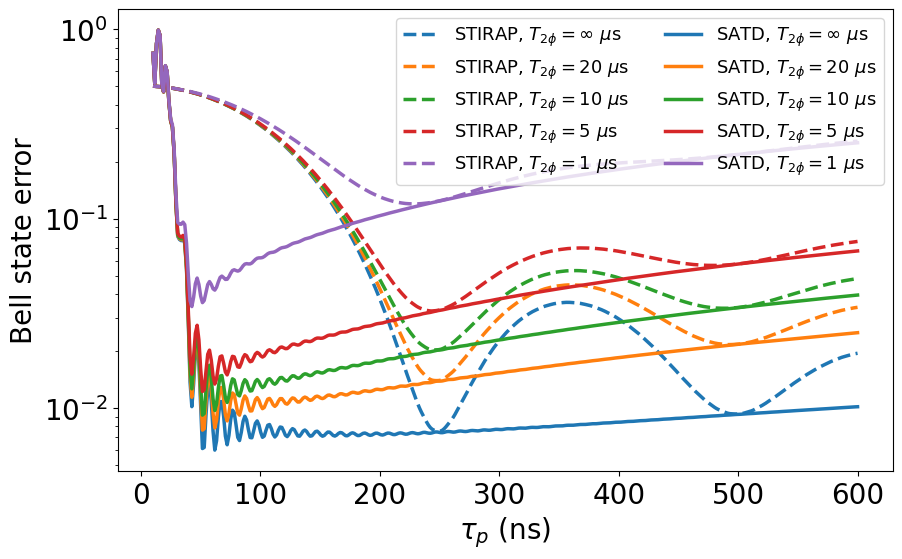} 
\caption{\textbf{Dependence of Bell state error on pure dephasing:} comparing STIRAP and SATD as a function of $\tau_p$ for various dephasing times $T_{2\phi,a}=T_{2\phi,a} \in \{\infty, 20, 10, 5, 1\} \ \mu$s. Other parameters are set as $T_{1a}=T_{1b}=100 \ \mu$s, $Q_c=10^{5}$, $g/2\pi=4$  MHz, and $\Delta_c/2\pi=100$ MHz. STIRAP and SATD results are shown with dashed and solid curves with the same colors for each $T_{2\phi}$ value.}
\label{fig:SATD-Bell deph}
\end{figure}

\begin{figure}[t!]
\centering
\includegraphics[scale=0.37]{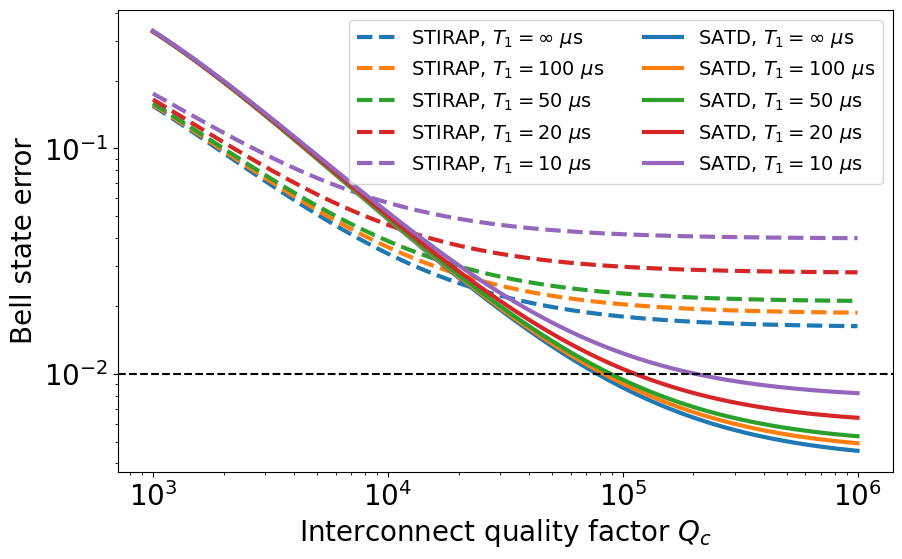} 
\caption{\textbf{Dependence of Bell state error on qubit/interconnect relaxation:} comparing STIRAP (dashed) and SATD (solid) results as a function of interconnect quality factor $Q_c$ for various qubit lifetimes of $T_{1a}=T_{1b}\in \{\infty, 100, 50, 20, 10\}$ $\mu$s, fixed $T_{2\phi,a}=T_{2\phi,b}=10$ $\mu$s and $g/2\pi=4$ MHz. Operation times for STIRAP (SATD) are picked according to the corresponding minimum in Fig.~\ref{fig:SATD-Bell deph} as $\tau_p=250$ ns ($\tau_p=51.5$ ns).}
\label{fig:SATD-Bell Relax}
\end{figure}

Figure~\ref{fig:SATD-STIRAP-SATD BELL comparison} shows a comparison between STIRAP and SATD, similar to that of Fig.~\ref{fig:SATD-STIRAP-SATD ST comparison} with zero relaxation and dephasing, for Bell state generation. Panel (a) shows that the SATD solution cancels out the non-adiabatic error entirely. The five mode simulation, however, manifests a constant floor for the error at sufficiently long times independent of $\tau_p$. The breakdown of interconnect mode populations in panels (b)--(c) reveals the source of this error as loss of qubit population to the odd modes which is almost equal between STIRAP and SATD. We find this adiabatic overlap of the supposedly dark state with the odd modes to scale approximately as $(g/\Delta_c)^2$. Here, we have picked a weak coupling of $g/2\pi=4$ MHz, which suppresses the overlap error down to $3.2\times 10^{-3}$. In this regime, SATD gives a substantial speedup, where the fastest operation times for STIRAP and SATD are approximately 250 ns and 86 ns, respectively.  Furthermore, Figs.~\ref{fig:SATD-Bell deph} and~\ref{fig:SATD-Bell Relax} characterize the sensitivity of Bell state generation error to pure dephasing and qubit/interconnect relaxation in a similar format as in Figs.~\ref{fig:SATD-ST deph}--\ref{fig:SATD-ST Relax} but for $g/2\pi = 4$ MHz. In this weak coupling limit, SATD leads to a noticeable reduction in error sensitivity to pure dephasing, where at $\tau_p \approx 51.5$ ns we find $\Delta E_{\text{deph}}^{\text{SATD}} (1 \ \mu s)= 2.98\times 10^{-2}$, compared to $\Delta E_{\text{deph}}^{\text{STIRAP}} (1 \ \mu s)= 1.164\times 10^{-1}$ at $\tau_p \approx 250$ ns. Moreover, based on Fig.~\ref{fig:SATD-Bell Relax}, SATD achieves improved error sensitivity with respect to both $Q_c$ and $T_1$, where $\Delta E_{\text{q-rel}}^{\text{SATD}} (10 \ \mu s)\approx 3.6\times 10^{-3}$ compared to $\Delta E_{\text{q-rel}}^{\text{STIRAP}}(10 \ \mu s)\approx 2.37\times 10^{-2}$ at sufficiently large $Q_c$.        

\subsection{Practical impacts of SATD on indirect remote two-qubit gate schemes}
\label{SubSec:Appl}

\begin{figure}[t!]
\centering
\includegraphics[scale=0.485]{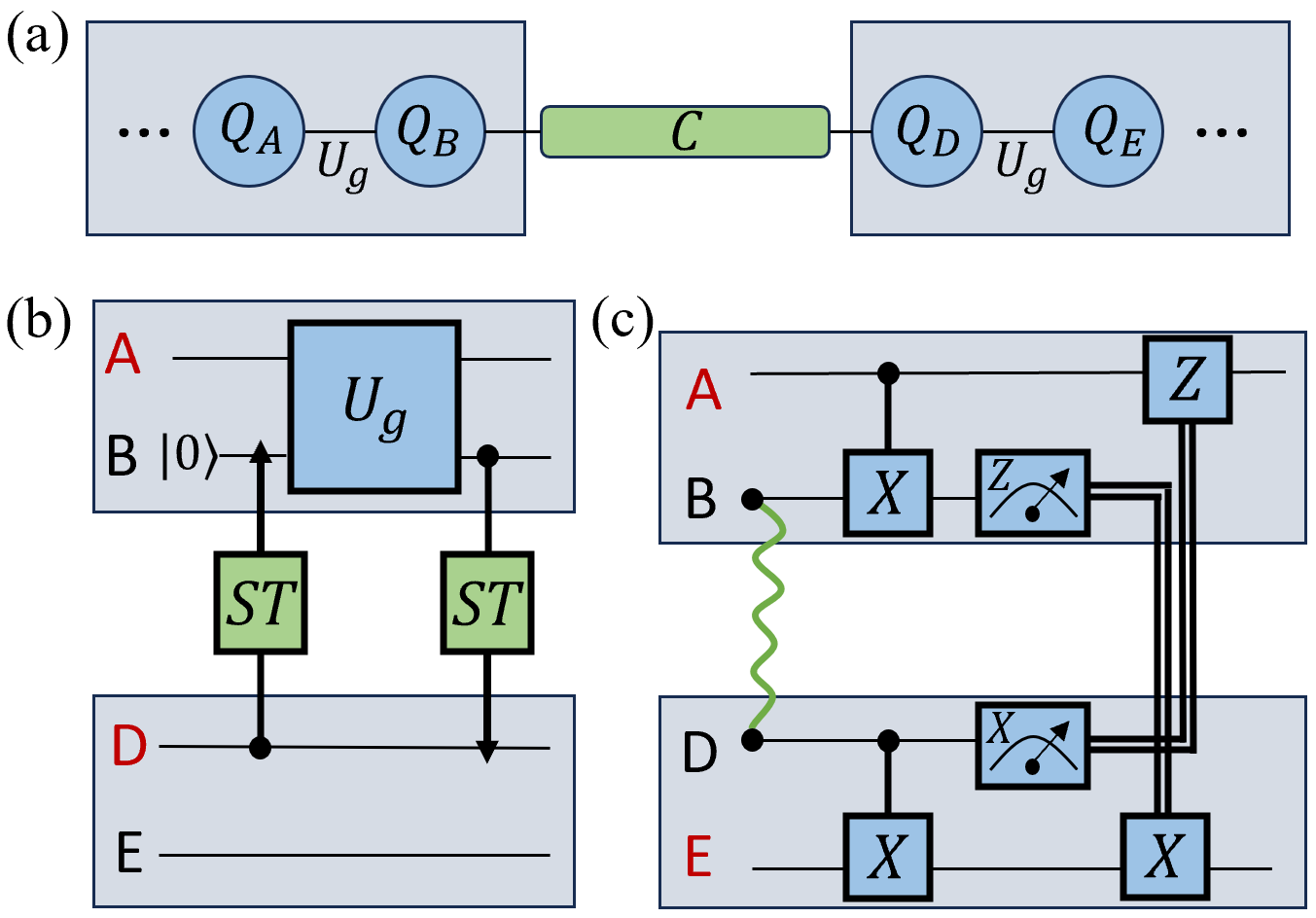} 
\caption{\textbf{Examples of remote \textit{indirect} two-qubit gate schemes:} (a) Schematics of two quantum modules coupled via an interconnect. (b) A protocol that combines an arbitrary local two-qubit gate $U_g$ between qubits A and B with two remote quantum state transfers between qubits B and D to ideally achieve the same, but remote, gate $U_g$ between qubits A and D. A similar two-qubit gate can be implemented between qubits B and E. (c) The CNOT gate teleportation protocol \cite{Eisert_Optimal_2000, Huang_Experimental_2004, Chou_Deterministic_2018, Wan_Quantum_2019} requires an initial Bell state, two local CNOTS, and two mid-circuit measurements and feedforward operations to achieve a remote CNOT between qubits A and E. The default entangled state in this protocol is $\ket{\Phi_+}\equiv (1/\sqrt{2})(\ket{00}+\ket{11})$ (squiggly line) which is equivalent to the state produced by STIRAP/SATD up to local $X$ and $Z$ operations (not shown for brevity).}
\label{fig:indirect_2Qgate_schemes}
\end{figure}

To put the SATD improvements into perspective, we revisit two indirect two-qubit gate schemes shown in Fig.~\ref{fig:indirect_2Qgate_schemes}. Imagine two QPU units with qubits A and B on the left, and D and E on the right. In each unit, we can perform a native two-qubit gate $\hat{U}_g$. However, assume that between the interface qubits B and D across the interconnect C, we can only perform quantum state transfer or generate entanglement. 

The first scheme in Fig.~\ref{fig:indirect_2Qgate_schemes}(b) allows the arbitrary native two-qubit gate $\hat{U}_g$ to act between an interface qubit and the qubit adjacent to the interface qubit on the other side. For instance, to perform a remote gate between qubits A and D we need to: (i) initialize qubit B in the ground state $\ket{0}$, (ii) transfer the state of D to B, (iii) perform the local native gate $\hat{U}_g$, and (iv) transfer the state of B back to D. In another words, $\hat{U}_{g,AD} = \hat{ST}_{B\rightarrow D}\hat{U}_{g,AB}\hat{ST}_{D\rightarrow B}$. A similar gate could be implemented between qubits B and E. Assuming sufficiently high fidelity for each individual operation, the average gate error \cite{Pedersen_Fidelity_2007} up to the leading order is roughly $\bar{E}_{g,AD} \approx \bar{E}_{U_g} + 2\bar{E}_{ST}$. Given the requirement for two state transfers, we expect the use of SATD to give noticeable improvement in both the gate speed and the average error. For instance, with $g/2\pi\approx 15$ MHz, Ref.~\cite{Chang_Remote_2020} calibrates a 130 ns state transfer using STIRAP, while with SATD we expect a $\mathcal{O}(50)$ ns transfer time for FSR of $\mathcal{O}(100)$ MHz. 

Figure~\ref{fig:indirect_2Qgate_schemes}(c) shows the well-known CNOT teleportation scheme \cite{Eisert_Optimal_2000, Huang_Experimental_2004, Chou_Deterministic_2018, Wan_Quantum_2019}. The protocol requires an initial entangled Bell pair between the interface qubits B and D, which can be prepared using STIRAP or SATD. Applying two local CNOT gates and two mid-circuit measurements and feedforward operations across the interconnect yields an effective CNOT gate between the outer qubits A and E. We expect the fidelity of the protocol to be mainly limited by the relatively long mid-circuit measurements and feedforward operations. The requirement for a weaker optimal $g/2\pi\approx 4$ MHz, however, makes the use of SATD more crucial, which can expedite the Bell generation substantially e.g. from 250 ns down to $\mathcal{O}(50)$ ns (see Fig.~\ref{fig:SATD-Bell deph}).


\section{Conclusion and outlook}
\label{Sec:Conc}

In this work, we promote the general application of shortcuts to adiabaticity methods \cite{Jarzynski_Generating_2013, Deffner_Classical_2014, Guery_Shortcuts_2019}, in particular SATD for STIRAP \cite{Baksic_Speeding_2016,Ribeiro_Accelerated_2019,Setiawan_Analytic_2021,Setiawan_Fast_2023}, in improving remote entanglement generation and quantum state transfer in multimode interconnects. Our results have applications to both long-range QPU-QPU and potential on-chip connections. Besides introducing new leakage and decay channels, we find the multimode nature of an interconnect violates the dark state symmetry required for adiabatic passage by an adiabatic overlap error with the odd modes that grows as $(g/\Delta_c)^2$, which impacts entanglement generation more strongly. This observation makes SATD a great fit in this multimode context, since due to its robustness against $g$ it allows for fast quantum operations at a sufficiently weak $g$ that suppresses the overlap error as well. For a meter-long interconnect with FSR of 100 MHz, we can calibrate $\mathcal{O}(50)$ ns quantum state transfer and Bell state with sub-percent error.   

We find the single-interconnect-mode SATD solutions \cite{Baksic_Speeding_2016,Ribeiro_Accelerated_2019,Setiawan_Analytic_2021,Setiawan_Fast_2023} to work approximately as intended in the weak $g$ limit such that only leakage in the resonant subspace (dark-bright transitions) is cancelled out, and the operation speed is limited by leakage to the adjacent modes, which is set by the interconnect FSR. A potential future research direction is expediting the operation even further by suppressing leakage to the off-resonant modes. One could ask whether precise or approximate SATD solutions exist in the multimode case, and whether they can be implemented via the same control knobs. Also, the single-mode SATD solutions can potentially serve as a reasonable initial guess for numerical optimal control techniques \cite{Khaneja_Optimal_2005, Koch_Quantum_2022} for further leakage improvement.        

\section{Acknowledgements}
\label{Sec:Acknow}
We appreciate helpful discussions with the IBM Quantum members Vikesh Siddhu, Theodore J. Yoder, Alireza Seif, Luke C. G. Govia, Muir Kumph, Jerry M. Chow, and Jay M. Gambetta. The authors acknowledge the IBM Research Cognitive Computing Cluster~(CCC) service for providing resources that have contributed to the research results reported within this paper.

\appendix
\section{Lindblad simulation}
\label{App:Lindblad}

For our numerical modeling of STIRAP and SATD, we run Lindbald simulations with two qubits and a finite number of interconnect modes. The Hamiltonian can be approximately described as
\begin{align}
\begin{split}
\hat{H}_s(t) & = \sum\limits_{q=a,b}\left[\omega_q(t) \hat{q}^{\dag}\hat{q}+\frac{1}{2}\alpha_q \hat{q}^{\dag}\hat{q}^{\dag}\hat{q}\hat{q}\right] + \sum\limits_{n-n_c=-N}^{N} \omega_{n} c_n^{\dag}\hat{c}_n \\
&+ \sum\limits_{n-n_c=-N}^{N} g_{an}(t) \left(\hat{a}\hat{c}_n^{\dag}+\hat{a}^{\dag}\hat{c}_n\right) \\
&+ \sum\limits_{n-n_c=-N}^{N} g_{bn}(t) (-1)^n\left(\hat{b}\hat{c}_n^{\dag}+\hat{b}^{\dag}\hat{c}_n\right) \;,
\end{split}
\label{Eq:Lindblad-Def of H}
\end{align}
with $\hat{a}$, $\hat{b}$ and $\hat{c}_n$ denoting the qubits and the $n^{th}$ interconnect modes, respectively. Moreover, $\omega$, $\alpha$ and $g$ represent mode frequency, anharmonicity and exchange interaction, respectively. The multimode interconnect is modeled as a set of $2N+1$ linear quantum harmonic oscillators as $\omega_{n} = \omega_{n_c} + n \Delta_c $ with the center frequency $\omega_{n_c}$ and FSR $\Delta_c$.

We account for various incoherent error sources such as qubit relaxation, cable relaxation, and qubit pure dephasing, by numerically solving the following Lindblad equation for the system density matrix $\hat{\rho}_s(t)$:  
\begin{align}
\begin{split}
\dot{\hat{\rho}}_s(t) &= - i [\hat{H}_s(t),\hat{\rho}_s(t)] + \sum\limits_{q=a,b} \frac{1}{T_{1q}}\mathcal{D}[\hat{q}]\hat{\rho}_s(t) \\
& + \sum\limits_{q=a,b} \frac{2}{T_{2\phi,q}}\mathcal{D}[\hat{q}^{\dag}\hat{q}]\hat{\rho}_s(t) + \sum\limits_{n} \kappa_n \mathcal{D}[\hat{c}_n]\hat{\rho}_s(t) \;,
\label{Eq:Lindblad-Lindblad eq.}
\end{split}
\end{align}
where $T_{1q}$ and $T_{2\phi,q}$ are the relaxation and pure dephasing times for qubit $q=a, b$, respectively, and $\kappa_n$ is the decay rate of the $n^{th}$ interconnect mode. Furthermore, $\mathcal{D}[\hat{C}]\hat{\rho}_s \equiv \hat{C}\hat{\rho}_s\hat{C}^{\dag} - (1/2)\{\hat{C}^{\dag}\hat{C},\hat{\rho}_s\}$ is the dissipator for the collapse operator $\hat{C}$. 

\begin{figure}[t!]
\centering
\includegraphics[scale=0.37]{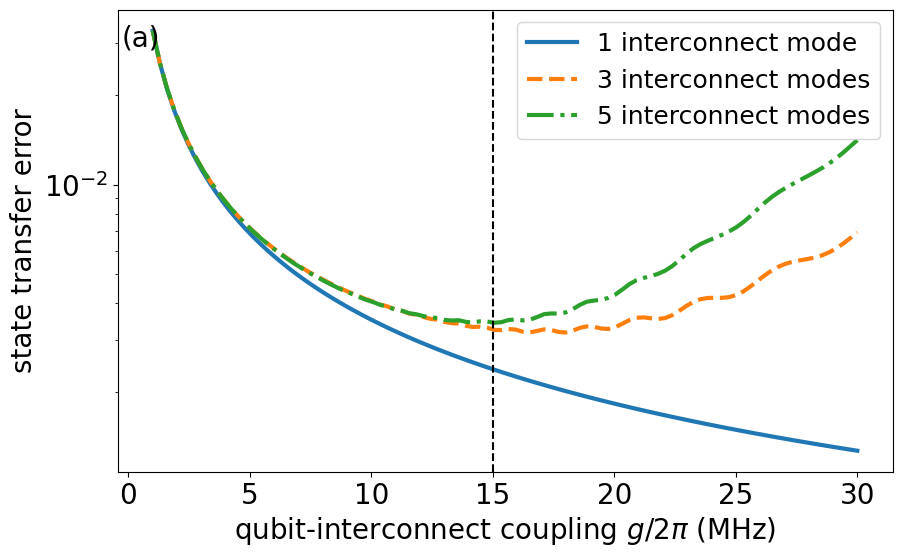}\\
\includegraphics[scale=0.37]{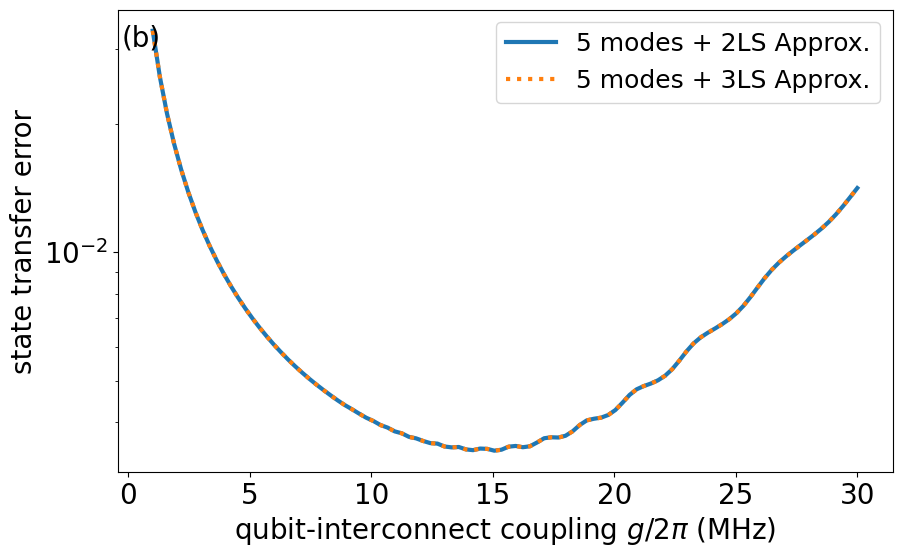}
\caption{\textbf{Numerical convergence test:} State transfer error for STIRAP when (a) including one, three and five interconnect modes, and (b) making two-level or three-level approximations for the qubits. Parameters are set similar to that of Fig.~\ref{fig:MMSTIRAP-ST Bell Error}(a) with $Q_c=10^5$. The vertical dashed line in panel (a) shows the largest coupling used in our simulations.}
\label{fig:Lindblad-ConvTest}
\end{figure}

A few remarks are in order. First, the qubit Hamiltonian is expressed as a multi-level Kerr oscillator. For the purpose of modeling state transfer and Bell state generation, however, the time-evolution is fairly accurately described by the one-excitation subspace. Therefore, the two-level approximation works well. Second, we have preformed RWA on the qubit-interconnect interactions given the experimentally relevant realizations where $g_{qc}/2\pi \approx \mathcal{O}(10)$ MHz, and $\omega_q/2\pi \approx \mathcal{O}(5)$ GHz. Third, the qubit-interconnect interaction rates $g_{qn}$ for $q=a,b$ in principle depend on the mode number approximately as $g_{qm}/g_{qn} \approx (\omega_{m}/\omega_{n})^{1/2}$ \cite{Houck_Controlling_2008, Malekakhlagh_Origin_2016, Gely_Convergence_2017, Malekakhlagh_Cutoff-Free_2017}. For a long coupler, however, qubits are resonant with a high-order interconnect mode, making the modal dependence of the interaction less pronounced. Fourth, the phase factor $(-1)^n$ for qubit $b$ interaction rate accounts for the opposite amplitude sign of even and odd spatial modes at the two ends of the interconnect.

For all results in the main text, we have accounted for five interconnect modes; one resonant with the qubits and two on each side. Even under two-level approximation for the qubits and the modes, this constitutes a large density matrix of dimension 128$\times$128 ($D = 2^7 = 128$). Figure~\ref{fig:Lindblad-ConvTest}(a) shows a convergence test of state transfer error with one, three and five interconnect modes. Generally speaking, stronger $g$ leads to more pronounced multimode effects involving further detuned modes. Our choice of five interconnect modes is a balance between simulation precision and speed for $g/2\pi$ of up to approximately 15 MHz used in the main text. Furthermore, since STIRAP is excitation preserving, higher-levels of the qubits would not impact the dynamics as shown in Fig.~\ref{fig:Lindblad-ConvTest}(b).

Lastly, we note that our numerical simulation of the Lindblad dynamics (\ref{Eq:Lindblad-Def of H})--(\ref{Eq:Lindblad-Lindblad eq.}) was performed using Qiskit Dynamics \cite{Puzzuoli_Algorithms_2023, Puzzuoli_Qiskit_2023}, along with standard Python scientific computing packages including \texttt{numpy} \cite{Harris_Array_2020} and \texttt{scipy} \cite{Virtanen_Scipy_2020}. We used the \texttt{DOP853} ODE solver with absolute and relative tolerances set to \texttt{atol=rtol=1e-10}. The simulations were parallelized over multiple CPU cores (up to 250) on IBM's Cognitive Computing Cluster. 

\section{Single-mode STIRAP}
\label{App:STIRAP}

Consider a resonant Lambda system with tunable interaction rates between the qubit states $q=a,b$ and coupler state $c$ as:
\begin{align}
\begin{split}
\hat{H}_{\text{STRP}}(t) & = \begin{bmatrix}
0 & g_{ac}(t) & 0\\
g_{ac}(t) & 0 & g_{bc}(t) \\
0 & g_{bc}(t) & 0 
\end{bmatrix}\\
& = \begin{bmatrix}
0 & g(t) \sin\theta(t) & 0\\
g(t)\sin\theta(t) & 0 & g(t)\cos\theta(t) \\
0 & g(t)\cos\theta(t) & 0 
\end{bmatrix} \;,
\end{split}
\label{Eq:STIRAP-Def of H_strp}
\end{align}
where in the second step we have re-expressed the interactions as $g(t)\equiv \sqrt {g_{ac}^2(t) +g_{bc}^2(t)}$ and $\tan \theta(t) \equiv g_{ac}(t)/g_{bc}(t)$. This resonant $\Lambda$ system contains two bright and one dark instantaneous eigenstates, where dark refers to no overlap with the intermediate interconnect state. Explicitly, Hamiltonian~(\ref{Eq:STIRAP-Def of H_strp}) can be diagonalized as
\begin{align}
\begin{split}
\hat{H}_{\text{INST}} & \equiv 	\hat{U}_{\text{INST}}\hat{H}_{\text{STRP}}\hat{U}_{\text{INST}}^{\dag} \\
& = \begin{bmatrix}
+g(t) & 0 & 0\\
0 & 0 & 0 \\
0 & 0 & -g(t)  
\end{bmatrix}
\end{split}\;,
\label{Eq:STIRAP-Def of H_inst}
\end{align}
via the unitary transformation
\begin{align}
\hat{U}_{\text{INST}}=\begin{bmatrix}
\sin\theta(t)/\sqrt{2} & 1/\sqrt{2} & \cos\theta(t)/\sqrt{2}\\
\cos\theta(t) & 0 & -\sin\theta(t) \\
\sin\theta(t)/\sqrt{2} & -1/\sqrt{2} & \cos\theta(t)/\sqrt{2} 
\end{bmatrix}\;,
\label{Eq:STIRAP-Def of U_inst}
\end{align}
where the rows of Eq.~(\ref{Eq:STIRAP-Def of U_inst}) represent the bright and dark eigenstates, having eigenergies $E_{B,\pm}(t) = \pm g(t)$ and $E_D(t) = 0$, respectively.

Under STIRAP, we adiabatically evolve the dark state $\ket{D(t)} = \cos\theta(t)\ket{1_a0_c0_b} - \sin\theta(t)\ket{0_a0_c1_b}$ of the system by sweeping the angle $\theta(t)$. Starting with $\theta(0) = 0$, we can create a Bell state or perform state transfer at final angle $\theta(\tau_p) = \pi/4$ and $\theta(\tau_p) = \pi/2$, respectively. Common control pulse shapes are
\begin{align}
& g_{ac}(t) = g \sin\theta(t) \;, 
\label{Eq:STIRAP-STIRAP gac}\\
& g_{bc}(t) = g \cos\theta(t) \;,
\label{Eq:STIRAP-STIRAP gbc}\\
& \theta(t) = \theta_p \frac{t}{\tau_p} \;.
\label{Eq:STIRAP-STIRAP theta}
\end{align}
which keeps the bright-dark transition frequency constant in time equal to $g$. The non-adiabatic error of STIRAP is explicitly found by the time-dependent transformation of the Schr\"odinger equation in the instantaneous frame as:
\begin{align}
\begin{split}
\hat{H}_{\text{NAD}} & \equiv i\dot {\hat{U}}_{\text{INST}} \hat{U}_{\text{INST}}^{\dag} \\
& = 
\begin{bmatrix}
 0 & +i\dot{\theta}(t) & 0\\
-i\dot{\theta}(t) & 0 & -i\dot{\theta}(t) \\
 0 & +i\dot{\theta}(t) & 0  
\end{bmatrix}
\end{split} \;,
\end{align}
which is responsible for dark-bright state transitions whose strength is determined by the STIRAP speed $\dot{\theta}(t)$. 

We can derive leading-order expressions for the dark-bright transitions using Magnus expansion \cite{Magnus_Exponential_1954, Blanes_Magnus_2009, Blanes_Pedagogical_2010}. Employing the control pulse shapes~(\ref{Eq:STIRAP-STIRAP gac})--(\ref{Eq:STIRAP-STIRAP gbc}), and in the frame rotating with the instantaneous Hamiltonian~(\ref{Eq:STIRAP-Def of H_inst}), the non-adiabatic Hamiltonian is transformed to 
\begin{align}
\begin{split}
\hat{\tilde{H}}_{\text{NAD}}(t) &= e^{i\hat{H}_{\text{INST}} t} \hat{H}_{\text{NAD}}(t) e^{-i\hat{H}_{\text{INST}} t}\\ 
&=\begin{bmatrix}
 0 & +i\dot{\theta}(t) e^{-i g t} & 0\\
-i\dot{\theta}(t) e^{+i g t} & 0 & -i\dot{\theta}(t)e^{+i g t} \\
 0 & +i\dot{\theta}(t)e^{-i g t} & 0  
\end{bmatrix}
\end{split} \;.
\end{align}
Up to the lowest order, the Magnus generator and the time-evolution operator are found as \cite{Blanes_Pedagogical_2010}:
\begin{align}
& \hat{\tilde{U}}_{\text{NAD}}(t,0)  = \hat{I} - i\hat{\tilde{G}}_1(t,0) + \mathcal{O} (\hat{\tilde{H}}_{\text{NAD}}^2(t)) \;,
\label{Eq:STIRAP-1st ord U_NAD}\\
&\hat{\tilde{G}}_1(t,0) \equiv \int_{0}^{t} dt' \hat{\tilde{H}}_{\text{NAD}}(t') \;. 
\label{Eq:STIRAP-1st ord G}
\end{align}
We finally compute the dark-bright transition probability up to the leading order as: 
\begin{align}
\begin{split}
P_{D \rightarrow B_{\pm}}(t,0) & \equiv \left|\bra{B_{\pm}(t)}\hat{\tilde{U}}_{\text{NAD}}(t,0)\ket{D(0)}\right|^2 \\
&\approx \left| \int_{0}^{t} dt'\, \dot{\theta}(t') e^{\mp ig t'}\right|^2\\
&+ \mathcal{O}(\hat{\tilde{H}}_{\text{NAD}}^2(t)) \;.
\label{Eq:STIRAP-1st ord P_{D->B}}
\end{split}
\end{align}
Based on Eq.(\ref{Eq:STIRAP-1st ord P_{D->B}}), for constant $\dot{\theta}(t) = \theta_p/\tau_p$, such a leakage can be minimized if the operation time satisfies $g\tau_p = 2n\pi$ for $n\in \mathbb{N}$, explaining the regular STIRAP lobes as e.g. in Fig.~\ref{fig:SATD-STIRAP-SATD ST comparison}(a).

\section{STIRAP considerations for a multimode interconnect}
\label{App:MMSTIRAP}

In a multimode setting, where $g$ is a non-negligible fraction of the interconnect's FSR $\Delta_c$, the off-resonant modes have a detrimental effect on the STIRAP protocol. Here, we provide a simplified argument on why using STIRAP for entanglement generation is more prone to multimode error compared to state transfer. 

Note the qubits also form off-resonant Lambda systems with the adjacent interconnect modes. To study the role of off-resonant modes, and the even-odd sign-dependent interactions, consider the simplest multimode extension of Eq.~(\ref{Eq:STIRAP-Def of H_strp}) as: 
\begin{align}
\begin{split}
\hat{H}_{\text{STRP}}^{\text{MM}}(t) & = \begin{bmatrix}
0 & g_{ac}(t) & g_{ac}(t) & g_{ac}(t) & 0\\
g_{ac}(t) & -\Delta_c & 0 & 0 & -g_{bc}(t) \\
g_{ac}(t) & 0 & 0 & 0 & g_{bc}(t) \\
g_{ac}(t) & 0 & 0 & \Delta_c & -g_{bc}(t) \\
0 & - g_{bc}(t) & g_{bc}(t) & -g_{bc}(t) & 0 
\end{bmatrix} \;,
\end{split}
\label{Eq:MMSTIRAP-Def of H_strp^off}
\end{align}
where we add two adjacent interconnect modes with detunings $\pm \Delta_c$. 

Due to the sign-dependent interactions, Hamiltonian~(\ref{Eq:MMSTIRAP-Def of H_strp^off}) only supports a pseudo-dark eigenstate of the form:  
\begin{align}
\ket{D_{\text{pseudo}}(t)} \propto 
\begin{bmatrix}
\cos[\theta(t)]\\
\frac{g(t)}{\Delta_c}\sin[2\theta(t)]\\
0\\
-\frac{g(t)}{\Delta_c}\sin[2\theta(t)]\\
-\sin[\theta(t)]
\end{bmatrix} \;,
\label{Eq:MMSTIRAP-Def of |D_pseudo(t)>}
\end{align}
where $g(t)\equiv \sqrt{g_{ac}^2(t)+g_{b	c}^2(t)}$ and $\tan\theta(t)\equiv g_{ac}(t)/g_{bc}(t)$. 

According to Eq.~(\ref{Eq:MMSTIRAP-Def of |D_pseudo(t)>}), for the case of alternating interaction sign, the pseudo dark eigenstate has a non-zero overlap of magnitude $|[g(t)/\Delta_c]\sin[2\theta(t)]|$ with the one-excitation subspace of the odd interconnect modes. For a hypothetical case of same-sign interaction, however, we find the original dark state is supported with zero overlap with all interconnect modes. First, we emphasize that this unwanted overlap is an \textit{adiabatic} error, which is independent of the STIRAP speed $\dot{\theta}(t)$, and can only be mitigated by weaker interaction $g$. Second, when sweeping the mixing angle from $\theta(0)=0$ to $\theta(\tau_p)=\pi/2$ for state transfer, the unwanted end overlap is zero given that $\sin[2\theta(\tau_p)]=\sin(\pi)=0$. This is, however, not the case for arbitrary entanglement generation and Bell state generation with $\theta(\tau_p)=\pi/4$ in particular, making it more susceptible to such an adiabatic error.   

\begin{figure}[t!]
\centering
\includegraphics[scale=0.37]{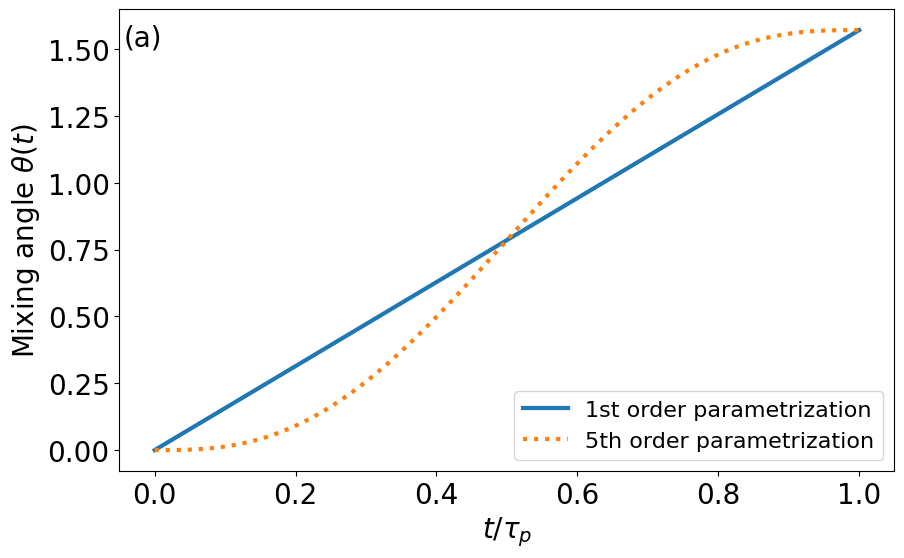} \\
\includegraphics[scale=0.37]{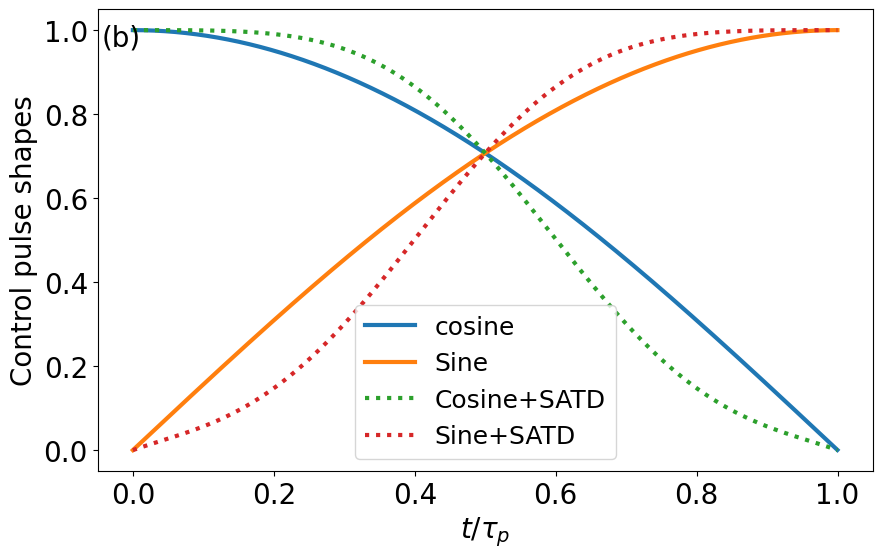} 
\caption{\textbf{STIRAP and SATD mixing angles and corresponding pulse shapes:} (a) First-order (STIRAP) and fifth-order (SATD) polynomial parametrization of mixing angle as in Eqs.~(\ref{Eq:STIRAP-STIRAP theta}) and~(\ref{Eq:SATD-SATD theta}) with $\theta_p = \pi/2$. (b) The corresponding STIRAP and SATD pulse shapes given in Eqs.~(\ref{Eq:STIRAP-STIRAP gac})--(\ref{Eq:STIRAP-STIRAP gbc}) and (\ref{Eq:SATD-SATD gac})--(\ref{Eq:SATD-SATD gbc}), respectively.}
\label{Fig:SATD-pulse shapes}
\end{figure}

\section{SATD correction for STIRAP}
\label{App:SATD}

Here, we review the derivation of SATD solutions for the STIRAP problem \cite{Baksic_Speeding_2016, Ribeiro_Accelerated_2019, Setiawan_Analytic_2021, Setiawan_Fast_2023}. Under the SATD method, we actively cancel out the non-adiabatic contribution by (i) correcting the controls, and (ii) dressing the adiabatic evolution path. Here, we review the derivation of a special SATD solution in which both the control and the dressing is along the $x$ direction: 
\begin{align}
&\hat{H}_{\text{CTRL}}(t) \equiv \hat{U}_{\text{INST}}^{\dag}(t) \left[h_x(t)\hat{M}_x\right]\hat{U}_{\text{INST}}(t) \;,
\label{Eq:SATD-Def of H_CNRL}\\
&\hat{V}(t)  \equiv \hat{R}_x[\mu(t)] = \exp[i\mu(t)\hat{M}_x] \;,
\label{Eq:SATD-Def of V}
\end{align}
where $h_x(t)$ and $\mu(t)$ are the x-control amplitude and x-dressing angle, respectively (to be determined), and $\hat{M}_k$ for $k=x, y, z$ is the spin-1 operator. 

We then solve for $h_x(t)$ and $\mu(t)$ such that in the frame dressed by $\hat{V}(t)$, given by
\begin{align}
\begin{split}
\hat{H}_{\text{DRS}}(t) &\equiv \hat{V}(t) \left[g\hat{M}_z(t) + \dot{\theta}(t)\hat{M}_y(t) \right. \\
& \left. + h_x(t)\hat{M}_x(t)\right] \hat{V}^{\dag}(t) + i\dot{\hat{V}}(t)\hat{V}^{\dag}(t) \;,
\end{split}
\end{align}
the off-diagonal non-adiabatic transitions are cancelled out at arbitrary time $t$. Enforcing the cancellation results in the following equations:
\begin{align}
\tan\mu(t) &= -\frac{\dot{\theta}(t)}{g} \;,\\
h_x(t) & = \dot{\mu}(t) \;,
\end{align}
with an explicit solution for $h_x(t)$ as:
\begin{align}
h_x(t) = - \frac{g\ddot{\theta}(t)}{g^2 + \dot{\theta}(t)^2} \;.
\end{align}
Transforming the corrected control back to the lab frame according to Eq.~(\ref{Eq:SATD-Def of H_CNRL}) one finds the SATD solutions as:
\begin{align}
& g_{ac}(t) = g \Big[\sin\theta(t) \underbrace{+\frac{\cos [\theta(t)] \ddot{\theta} (t)}{g^2+\dot{\theta}(t)^2}}_{\text{SATD correction}}\Big] \;,
\label{Eq:SATD-SATD gac}\\
& g_{bc}(t) = g \Big[\cos\theta(t) \underbrace{-\frac{\sin [\theta(t)] \ddot{\theta} (t)}{g^2+\dot{\theta}(t)^2}}_{\text{SATD correction}}\Big] \;.
\label{Eq:SATD-SATD gbc}
\end{align}
To ensure that the initial and final points of the adiabatic evolution remains unchanged we further enforce $\dot{\theta}(t)|_{t=0,\tau_p}=\ddot{\theta}(t)|_{t=0,\tau_p}=0$, in addition to $\theta(0)=0$ and $\theta(\tau_p)=\theta_p$. The lowest-order polynomial satisfying these conditions is then found as:  
\begin{align}
\theta^{(5)}(t)= \theta_p \left[6\left(\frac{t}{\tau_p}\right)^5-15\left(\frac{t}{\tau_p}\right)^4+10\left(\frac{t}{\tau_p}\right)^3\right] \;.
\label{Eq:SATD-SATD theta}
\end{align}
Figure~\ref{Fig:SATD-pulse shapes} shows a comparison between the modified SATD control pulse shapes~(\ref{Eq:SATD-SATD gac})--(\ref{Eq:SATD-SATD theta}) and the standard STIRAP control in Eqs.~(\ref{Eq:STIRAP-STIRAP gac})--(\ref{Eq:STIRAP-STIRAP theta}).

\bibliographystyle{unsrt}
\bibliography{LCoupler_SATD_Bibliography}

\end{document}